\documentclass[prb,twocolumn]{revtex4}%

\usepackage{graphicx}
\usepackage{amsmath}
\usepackage{ulem}
\usepackage{color}

\newcommand{\be}{\begin{equation}}
\newcommand{\ee}{\end{equation}}
\newcommand{\bea}{\begin{eqnarray*}}
\newcommand{\eea}{\end{eqnarray*}}
\newcommand{\bean}{\begin{eqnarray}}
\newcommand{\eean}{\end{eqnarray}}

\begin{document}

\draft
\title{\bf  Contact effects on thermoelectric properties of  textured graphene nanoribbons }

\author{David M T Kuo$^{1}$ and Yia-Chung Chang$^{2,3}$}

\address{$^{1}$Department of Electrical Engineering and Department of Physics, National Central
University, Chungli, 320 Taiwan}

\address{$^{2}$Research Center for Applied Sciences, Academic Sinica,
Taipei, 11529 Taiwan}

\affiliation{$^3$ Department of Physics, National Cheng Kung
University, Tainan, 701 Taiwan}

\date{\today}

\begin{abstract}
Transport and thermoelectric properties of finite textured
graphene nanoribbons (t-GNRs) connected to electrodes with various
coupling strengths are theoretically studied in the framework of
the tight-binding model and Green's function approach. Due to
quantum constriction induced by the indented edges, such t-GNRs
behave like serially-coupled graphene quantum dots (SGQDs). These
types of SGQDs can be formed by tailoring zigzag GNRs (ZGNRs) or
armchair GNRs (AGNRs). Their bandwidths and gaps can be engineered
by varying the size of the quantum dot and the neck width at
indented edges. Effects of defects and contact junction on
electrical conductance, Seebeck coefficient and electron thermal
conductance of t-GNRs are calculated. When a defect occurs in the
interior site of textured ZGNRs (t-ZGNRs), the maximum power
factor within the central gap or near the band edges is found to
be insensitive to the defect scattering. Furthermore, we found
that SGQDs formed by t-ZGNRs have significantly better electrical
power outputs than those of textured ANGRs due to the improved
functional shape of the transmission coefficient in t-ZGNRs. With
a proper design of contact the maximum power factor ( figure of
merit) of t-ZGNRs could reach $90\%$ ($95\%$) of the theoretical
limit.
\end{abstract}

\maketitle

\section{Introduction}

Due to  global warming, the  Kyoto protocol aiming to reduce
$CO_2$ emissions was proposed in 1997. Since then, renewable
energies including solar, wind, rain, tides, and geothermal heat
are topics of tremendous scientific interest [\onlinecite{ChenG}].
Thermoelectric devices can be used as power generators and
refrigerators. The electrical power based on thermoelectric
effects is one of the most important type of green energies
[\onlinecite{ChenG}-\onlinecite{Heremans}]. Because of the
inherent physics of thermoelectric effect, thermoelectric devices
can fully avoid $CO_2$, hydrofluorocarbons, and perfluorocarbons
emissions. In addition, thermoelectric devices can sustain long
operation time and avoid mechanical noises.

Recently, energy harvesting applications in nanoscale systems have
attracted considerable
attention[\onlinecite{Nika}-\onlinecite{Luo}]. How to obtain the
maximum thermoelectric efficiency of heat engines with optimized
electrical power output has been a key issue
[\onlinecite{Whitney}-\onlinecite{Luo}]. To improve the
performance of a heat engine it is preferable to have the electron
transport in ballistic regime and phonon transport in the
diffusive scattering regime. Therefore, a high-performance
thermoelectric device needs to provide a channel length shorter
than the electron mean free path ($\lambda_e$) but much longer
than the phonon mean free path ($\lambda_{ph}$) in order to reduce
the ratio of phonon thermal conductance ($\kappa_{ph}$) to
electron thermal conductance ($\kappa_{e}$) [\onlinecite{Xu}]. It
was pointed out in ref.~[\onlinecite{Whitney}] that a Carnot heat
engine favors the electron transport in an energy range where the
transmission coefficient has a steep change with respect to
energy, e.g. with a square-form (SF). Up to now it remains unclear
how to realize a SF transmission coefficient in realistic
thermoelectric devices with a short channel length between thermal
contacts, [\onlinecite{Darancet},\onlinecite{ChenIJ}].
%(L_c <\lambda_e

Considerable scientific efforts paved the way to answer such an
intriguing problem. Hicks and Dresselhaus theoretically
demonstrated that the thermoelectric performance can be
significantly enhanced in one-dimensional (1D) nanowires due to
the reduced phonon thermal conductance  and the enhanced Seebeck
coefficient ($S$) [\onlinecite{Hicks},\onlinecite{Dresselhaus}].
According to theoretical modeling, $\kappa_{ph}$ of 1D silicon
quantum dot (QD) superlattices can be reduced by one order of
magnitude in comparison with 1D nanowires
[\onlinecite{Nika},\onlinecite{HuM}]. Due to limitations in
technology the dot-size fluctuation in the 1D silicon quantum-dot
(QD) arrays still remains a serious issue.
[\onlinecite{Kagan},\onlinecite{Lawrie}]

The discovery of graphene in 2004 opened the door for realizing 1D
nanowires with small cross-section to degrade $\kappa_{ph}$
[\onlinecite{Novoselovks}], since one can fabricate graphene
nanoribbons (GNRs) with atomic precision via the bottom-up
approach[\onlinecite{Cai}]. This approach has been successfully
applied to build more complex systems, such as armchair GNRs with
periodically corrugated edges, called  as textured AGNRs (t-AGNRs)
here.[\onlinecite{ChenYC}-\onlinecite{DJRizzo}] The scanning
tunneling microscopy (STM) spectra of serially-coupled graphene
quantum dots (SGQDs) synthesized by t-AGNRs were reported
experimentally [\onlinecite{DJRizzo}]. Novel graphene-based
electronic devices have also focused on AGRNs and t-AGNRs
[\onlinecite{Llinas}]. Recently, the existence of edge states with
topological protection in textured zigzag GNRs (t-ZGNRs) has been
demonstrated theoretically [\onlinecite{Chou}].

Several experimental studies of thermoelectric properties of
graphene-related materials have been reported in recent years
[\onlinecite{ZuevYM}-\onlinecite{WangYY}]. Nevertheless, there is
a paucity of studies to consider the contact effect on
thermoelectric properties of t-AGNRs and
t-ZGNRs[\onlinecite{HeJ}]. In this work, our goal is to optimize
the transmission coefficient of t-GNRs with length shorter than
$\lambda_e$, but larger than $\lambda_{ph}$ under different
coupling strengths with the electrical contact. We found that
minibands can be formed in these t-GNRs, while their energy gaps
can be tuned by varying the size of GQDs and the inter-dot
coupling strength.Furthermore, the calculated power factor is very
robust against scattering from defects occurring inside the
interior of these t-ZGNRs. The transmission coefficients through
the minibands of t-ZGNRs under optimized tunneling rates provide
desired characteristics showing steep change with respect to
energy near the central gap that mimics the theoretical limit
obtained by using a SF transmission coefficient. Due to this
feature, our calculations reveal that the thermoelectric
performance of t-ZGNRs can be significantly better than the
t-AGNRs. The maximum power factor (figure of merit) of t-ZGNRs
could reach $90\%$ ($95\%$) of that obtained with the SF
transmission coefficient. Therefore, the t-ZGNR is a promising
candidate for applications in nanoscale energy harvesting.

\begin{figure}[h]
\centering
\includegraphics[trim=2.5cm 0cm 2.5cm 0cm,clip,angle=0,scale=0.3]{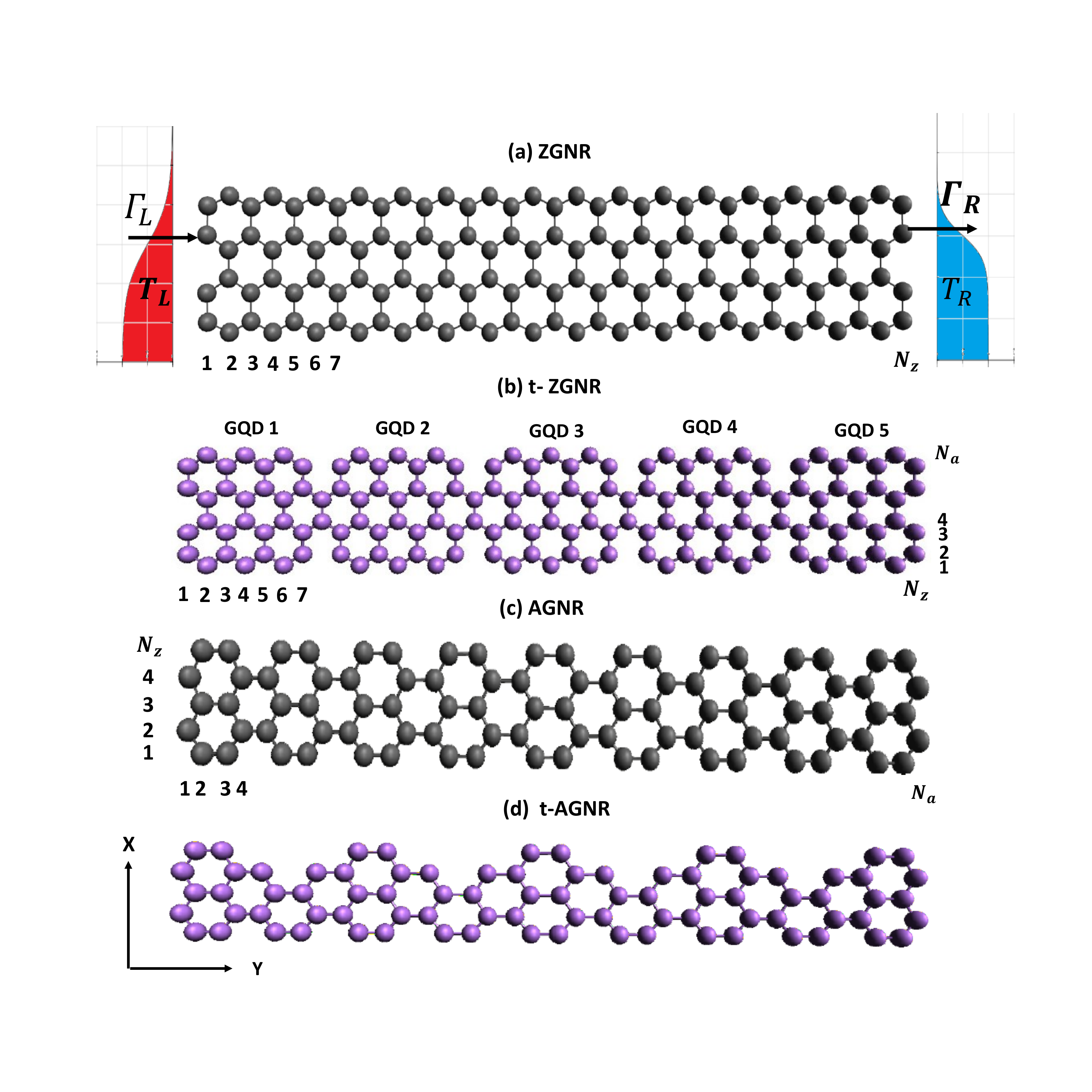}
\caption{Schematic diagrams of (a) ZGNR, (b) textured-ZGNR
(t-ZGNR), (c) AGNR, and (d) t-AGNR. In (a), we also show the
electrodes with  $\Gamma_{L}$ ($\Gamma_R$) denoting the tunneling
rate of the electrons between the left (right) electrode and the
leftmost (rightmost) atoms of the ZGNR. $T_L$ and $T_R$ denote the
temperature of the left ($L$) and the right ($R$) electrodes,
respectively. Note that  t-ZGNRs can be formed by periodically
removing some carbon atoms on the zigzag edges of ZGNRs. In (b)
the t-ZGNR consists of GQDs with size characterized by
$(N_a,N_z)=(8,7)$. In (c) the AGNR has a length $N_a=36$ and width
$N_z=5$. In (d), the width and length of the unit
cell for the t-AGNR superlattice are characterized by $N_z=5$ and
$N_a=8$, respectively.}
\end{figure}

\section{Calculation method}
To model the thermoelectric properties of t-GNRs connected to the
electrodes, it is a good approximation to employ a tight-binding
model with one $p_z$ orbital per atomic site to describe the
electronic states.[\onlinecite{Nakada}-\onlinecite{Sevincli}]
The Hamiltonian of the nano-junction system depicted in Fig.~1, including four different GNR structures, can be written as $H=H_0+H_{GNR}$
[\onlinecite{Haug}], where
\begin{small}
\begin{eqnarray}
H_0& = &\sum_{k} \epsilon_k a^{\dagger}_{k}a_{k}+
\sum_{k} \epsilon_k b^{\dagger}_{k}b_{k}\\
\nonumber &+&\sum_{\ell}\sum_{k}
V^L_{k,\ell,j}d^{\dagger}_{\ell,j}a_{k}
+\sum_{\ell}\sum_{k}V^R_{k,\ell,j}d^{\dagger}_{\ell,j}b_{k} + h.c.
\end{eqnarray}
\end{small}
The first two terms of Eq.~(1) describe the free electrons in the
left ($L$) and right ($R$) electrodes. $a^{\dagger}_{k}$
($b^{\dagger}_{k}$) creates an electron with wave number $k$ and
energy $\epsilon_k$ in the left (right) electrode.
$V^L_{k,\ell,j=1}$ ($V^R_{k,\ell,j=N_z(N_a)}$) describes the
coupling between the left (right) lead with its adjacent atom in
the $\ell$-th row.
\begin{small}
\begin{eqnarray}
H_{GNR}&= &\sum_{\ell,j} E_{\ell,j} d^{\dagger}_{\ell,j}d_{\ell,j}\\
\nonumber&-& \sum_{\ell,j}\sum_{\ell',j'} t_{(\ell,j),(\ell', j')}
d^{\dagger}_{\ell,j} d_{\ell',j'} + h.c,
\end{eqnarray}
\end{small}
where { $E_{\ell,j}$} is the on-site energy for the $p_z$ orbital
in the ${\ell}$-th row and $j$-th column. Here, the spin-orbit
interaction is neglected. $d^{\dagger}_{\ell,j} (d_{\ell,j})$
creates (destroys) one electron at the atom site labeled by
($\ell$,$j$) where $\ell$ and $j$, respectively are the row and
column indices as illustrated in Fig.~1. $t_{(\ell,j),(\ell',
j')}$ describes the electron hopping energy from site ($\ell$,$j$)
to site ($\ell'$,$j'$). The tight-binding parameters used for GNRs
is $E_{\ell,j}=0$ for the on-site energy and
$t_{(\ell,j),(\ell',j')}=t_{pp\pi}=2.7$ eV for nearest-neighbor
hopping strength.

To study the transport properties of a GNR junction connected to
electrodes, it is convenient to use the Keldysh Green's function
technique [\onlinecite{Haug}]. Electron and heat currents leaving
the electrodes can be expressed as
\begin{eqnarray}
J &=&\frac{2e}{h}\int {d\varepsilon}~ {\cal
T}_{LR}(\varepsilon)[f_L(\varepsilon)-f_R(\varepsilon)],
\end{eqnarray}
and
\begin{equation}
Q_{e,L(R)}=\frac{\pm 2}{h}\int {d\varepsilon}~ {\cal
T}_{LR}(\varepsilon)(\varepsilon-\mu_{L(R)})[f_L(\varepsilon)-f_R(\varepsilon)]
\end{equation}
where
$f_{\alpha}(\varepsilon)=1/\{\exp[(\varepsilon-\mu_{\alpha})/k_BT_{\alpha}]+1\}$
denotes the Fermi distribution function for the $\alpha$-th
electrode, where $\mu_\alpha$  and $T_{\alpha}$ are the chemical
potential and the temperature of the $\alpha$ electrode. $e$, $h$,
and $k_B$ denote the electron charge, the Planck's constant, and
the Boltzmann constant, respectively. ${\cal T}_{LR}(\varepsilon)$
denotes the transmission coefficient of a  GNR connected to
electrodes, which can be solved by the formula $ {\cal
T}_{LR}(\varepsilon)=4Tr[\Gamma_{L}(\varepsilon)
G^{r}(\varepsilon)\Gamma_{R}(\varepsilon)G^{a}(\varepsilon)]$
[\onlinecite{Kuo4},\onlinecite{Phung}], where
$\Gamma_{L}(\varepsilon)$ and $\Gamma_{R}(\varepsilon)$ denote the
tunneling rate (in energy units) at the left and right leads, and
{${G}^{r}(\varepsilon)$ and ${G}^{a}(\varepsilon)$ are the
retarded and advanced Green's functions of the GNR, respectively.
The tunneling rates are described by the imaginary part of
self-energy correction on the interface atoms of the GNR due to
the coupling with nearby atoms in the adjacent electrodes, i.e.
$\Gamma_{L(R)}(\varepsilon)=-Im(\Sigma^r_{L(R)}(\varepsilon)$).
Such a coupling depends on the contact quality with the
electrodes, which is characterized by the interaction strength
$V^L_{k,\ell,j=1} (V^R_{k,\ell,j=N_z(N_a)})$) with the left
(right) lead.} Here, we have adopted energy-independent tunneling
rates $\Gamma_{L(R)}(\varepsilon) =\Gamma_{L(R)}$ which is
reasonable in the wide-band limit for the electrodes
[\onlinecite{Kuo4}].Note that $\Gamma_{\alpha}$ and Green's
functions are matrices in the basis of tight-binding orbitals.
$\Gamma_{\alpha}$ for the boundary atoms have diagonal entries
given by the same constant $\Gamma_t$. Because of the line contact
as illustrated in Fig. 1, the contact resistance can be much
smaller than that of surface contact [\onlinecite{ChenRS}].
Meanwhile, the variation of tunneling rates could reveal different
contact properties such as the Schottky barrier or ohmic contact
[\onlinecite{Mahan}].

In the linear response regime, the electrical conductance ($G_e$),
Seebeck coefficient ($S$) and electron thermal conductance
($\kappa_e$) are given by $G_e=e^2{\cal L}_{0}$, $S=-{\cal
L}_{1}/(eT{\cal L}_{0})$ and $\kappa_e=\frac{1}{T}({\cal
L}_2-{\cal L}^2_1/{\cal L}_0)$ with ${\cal L}_n$ ($=0,1,2$)
defined as
\begin{equation}
{\cal L}_n=\frac{2}{h}\int d\varepsilon~ {\cal
T}_{LR}(\varepsilon)(\varepsilon-\mu)^n\frac{\partial
f(\varepsilon)}{\partial \mu}.
\end{equation}
Here $f(\varepsilon)=1/(exp^{(\varepsilon-\mu)/k_BT}+1)$ is the
Fermi distribution function of electrodes at equilibrium
temperature $T$ and chemical potential $\mu$.  As seen in Eq. (5),
the transmission coefficient ${\cal T}_{LR}(\varepsilon)$ plays a
significant role for electron transport between the left($L$) and
right ($R$) electrodes. At zero temperature, the electrical
conductance is given by $G_e(\mu)=\frac{2e^2}{h}{\cal
T}_{LR}(\mu)$. In the current study, only even numbers of $N_a$
are considered to avoid unwanted dangling-bond states. We have
developed an efficient computation method that allows us to
calculate the Green's functions of large-sized quantum structures.
For the current study, textured ZGNRs and AGNRs with lengths up to
$14 nm$ are considered.

\section{Results and discussion}

\subsection{Graphene nanoribbons}
{Topological states offer promising applications in electronics
and optoelectronics, owing to their robustness in transport
characteristics against defect scattering. Many studies have
confirmed that 2D and 1D topological states (TSs) exist in certain
material structures [\onlinecite{ChangYC}-\onlinecite{Xian}]. For
instance, there exists a 2D topologically protected interface
state in HgTe/CdTe superlattices [\onlinecite{ChangYC,ZhangSC}].
One-dimensional (1D) TSs were theoretically predicted to exist in
square selenene and tellurene [\onlinecite{Xian}]. Recently,
zero-dimensional (0D) TSs of finite-size GNRs have been
extensively studied [\onlinecite{ChenYC}-\onlinecite{DJRizzo}]
because the 0D TS offers more flexibility in the design of
complicated electronic circuits [\onlinecite{Yan}]. Before
illustrating electron coherent transport through SGQDs formed by
t-ZGNRs, we first examine the characteristics of 0D TSs of
finite-size GNRs by calculating their transmission coefficient,
${\cal T}_{LR}(\varepsilon)$.}

Figure~2 shows the calculated electron conductance, $G_e$ at
$k_BT=0$ as a function of $\mu$ for various $N_a$ with $N_z=7$. We
note that there are two zigzag edge states localized at the top
and bottom sides of the GQDs depicted in Fig.~1(a). For $N_a=40$,
their wave functions are well separated along the armchair
direction [\onlinecite {Wakabayashi}]. The electrical conductance
spectrum shows that $G_e=2G_0$ at $\mu=0$ and $G_e=G_0$ for other
electronic states, where $G_0=\frac{2e^2}{h}$ is the quantum
conductance. Such zero energy modes were observed experimentally
by STM [\onlinecite{DRizzo}]. When two electronic states are
closely spaced {(with energy separation less than the broadening),
$G_e$ can become large than $G_0$. As $N_a$ decreases, $G_e$ for
the zero-energy mode is split into two peaks corresponding to the
bonding and antibonding states of coupled zigzag edge states, as
seen in Fig.~2c. When $N_a=12$, these two peaks are well
separated. Here, $\epsilon_{HO}$ and $\epsilon_{LU}$ denote the
energy levels of the} highest occupied molecular orbital (HOMO)
and the lowest unoccupied molecular orbital (LUMO), respectively.
Due to the short decay lengths along the armchair edge directions
for zigzag edges states, ${\cal T}_{LR}(\varepsilon)$ of
$\Sigma_0$ also depends on {whether the zigzag or armchair edges
are} coupled to the electrodes. In Ref.~[\onlinecite{MahanG}] it
is proposed that a single QD can be used to realize a Carnot heat
engine. However, the channel length for $N_z=7$ ($L_z=0.738~nm$)
is too small to avoid the serious degradation of the figure of
merit due to the phonon heat conductivity, $\kappa_{ph}$. On the
other hand, for the large $N_z$ case ($N_z\gg N_a$), a finite GNR
shows metallic behavior, leading to unfavorable thermoelectric
properties. To maintain a sizable gap ($~10k_BT$) {around the
charge neutrality point (CNP)} while keeping $\kappa_{ph}$ small
enough to preserve decent figure of merit
($ZT=\frac{S^2G_eT}{\kappa_e+\kappa_{ph}}$), the t-ZGNR becomes a
suitable candidate as we shall discuss below.

\begin{figure}[h]
\centering
\includegraphics[angle=0,scale=0.3]{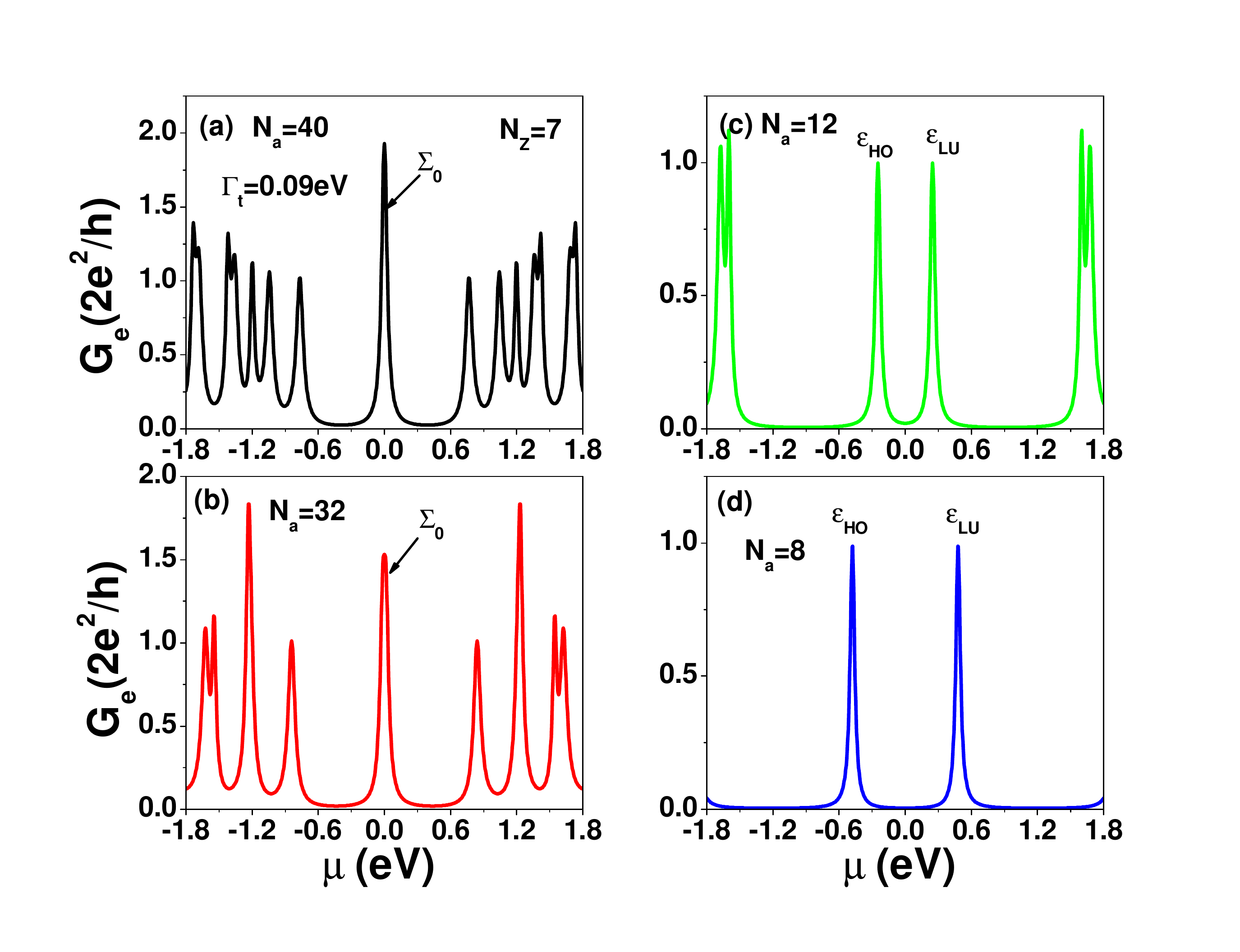}
\caption{Electrical conductance, $G_e$ of a finite GNR with
armchair edges coupled to the electrodes at $k_BT=0$ as a function
of the chemical potential, $\mu$ for various widths with
$N_a=40,32,12,$ and 8 with $N_z$ fixed at 7. We have adopted
electron tunneling rates $\Gamma_t=90$ meV.}
\end{figure}

\subsection{ SGQDs formed by textured ZGNRs}
For energy harvesting applications at room temperature, we need to
design a ${\cal T}_{LR}(\varepsilon)$ spectrum with a square shape
near the central gap to obtain optimized $ZT$ and electrical power
output [\onlinecite{Whitney}]. Let's consider an SGQD formed by
t-ZGNR. {A t-ZGNR can be realized by tailoring GNR with periodic
indentation on the zigzag edges such as the structure shown in
Fig.1(b). This t-ZGNR consists of GQDs with size characterized by
$N_a=12$ and $N_z=7$. A single GQD of this size has resonance
energies $\varepsilon_{LU(HO)}=\pm 0.247eV$ near the CNP, as shown
in Fig. 2(c). For two coupled GQDs (2GQDs), there is one satellite
peak on the left (right) side of the $\varepsilon_{HO(LU)}$ peak
(see Fig. 3(b)). For six coupled GQDs (6GQDs), we have five
additional peaks with $\varepsilon_{e(h),1}=\pm 0.3105eV$,
$\varepsilon_{e(h),2}=\pm 0.468eV$, $\varepsilon_{e(h),3}=\pm
0.6705eV$,$\varepsilon_{e(h),4}=\pm 0.882eV$ and
$\varepsilon_{e(h),5}=\pm 1.0665eV$ on the right (left) side of
$\varepsilon_{LU(HO)}=\pm 0.247eV$ (see Fig. 3(f)). We note that
$G_e(\mu)$ is fully suppressed for $\mu$ between
$\varepsilon_{HO}$ and $\varepsilon_{LU}$. Therefore, SGQDs formed
by t-ZGNRs function as filters to block} electrons with energies
near the CNP, which is different from that of serially coupled
antidots realized by ZGNRs with nanopores [\onlinecite{ChangPH}].
The separation between peaks is inhomogeneous. In addition, the
$\varepsilon_{HO}$ and $\varepsilon_{LU}$ peaks become sharper
with increasing GQD number. These features are attributed to a
special parity symmetry in the supercell of ZGNRs. Due to the
coupling of transverse and longitudinal wave numbers in ZGNRs, the
density distributions of electronic state near CNP are
inhomogeneous [\onlinecite{Wakabayashi2}]. AGNRs do not have such
a parity in the supercell. The feature provides a significant
effect on the transmission coefficient of SGQDs formed by t-ZGNRs.
To further illustrate the characteristics of the edge states we
show the charge densities of states with $\varepsilon_{LU}$,
$\varepsilon_{e,1}$ and $\varepsilon_{e,2}$ for the case of 6GQDs
 in Fig. A.1 of the appendix.

\begin{figure}[h]
\centering
\includegraphics[angle=0,scale=0.3]{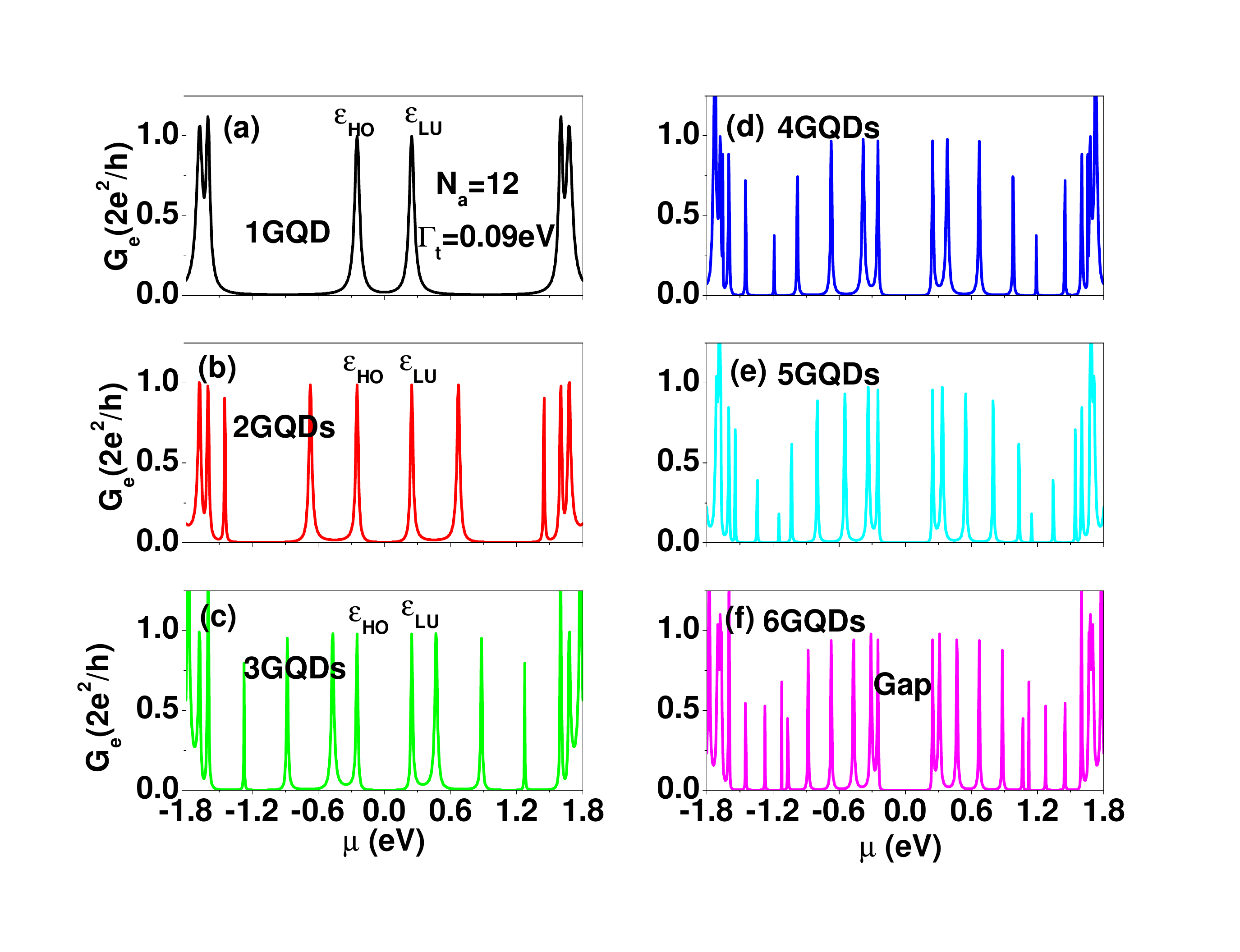}
\caption{Electrical conductance, $G_e$ as a function of $\mu$ for
SGQDs with various of GQD numbers. A SGQD consists of GQDs with
the size characterized by $N_a=12$ and $N_z=7$. Other physical
parameters correspond to those of Fig.~2(c).}
\end{figure}

{Previous theoretical studies have demonstrated that edge defects
can significantly reduce the electron quantum conductance of ZGNRs
[\onlinecite{Areshkin},\onlinecite{Martins}]. Here, we investigate
how defects (either at the edge or in the interior region)
influence the thermoelectric characteristics of t-ZGNRs by
introduce energy shifts $\Delta_{\ell,j}$ on designated defect
sites. $\Delta_{\ell,j}$ could be positive or negative, which
depends on the kind of defects [\onlinecite{Martins}]. To model a
vacancy within a tight binding model, one typically takes
$\Delta_{\ell,j} \rightarrow \infty$. The larger orbital-energy
shift, the stronger effect on the electrical conductance can be
seen [\onlinecite{LiTC}]. Here, we consider the case of a positive
and large $\Delta$ to investigate the effects of defects on the
electron transport of t-ZGNRs. We calculate $G_e$, $\kappa_e$, $S$
and power factor $PF=S^2 G_e$ for different defect locations for
the case of 15 GQDs ($L_z=14.5$nm or $N_z=119$) with
$\Gamma_t=0.54eV$ and show the results in Fig. 4.

As seen in Fig.~4(a) and 4(b), the size of the central gap and the
shape of $G_e$ and  $\kappa_e$ spectra (green lines) are changed
only slightly when the defects are located in the interior region.
However the $G_e$ and $\kappa_e$ values near the central gap
become seriously suppressed (as shown by red lines) when a single
defect is located at a zigzag edge site labeled by $(1,4)$, where
the charge density of the $\varepsilon_{LU(HO)}$ electronic state
is peaked. As seen in Fig.~4(c) the antisymmetric pattern of the
Seebeck coefficient $S(\mu)$ (with respect to the sign change of
$\mu$) due to the electron-hole symmetry is distorted in the
presence of defects. The peak value of the power factor
$PF=S^2G_e$ near the central gap is seriously serious reduced when
defects occur on the edge, but much less affected by defects in
the interior region as illustrated in Fig.~4(d).} Here and
henceforth, $\kappa_e$ is in units of $\kappa_0=0.62nW/K$,  $S$ is
expressed in units of $k_B/e=86.25\mu V/K$, and the power factor
($PF$) in units of $2k^2_B/h=0.575pW/K^2$. It is remarkable to see
that $PF$ is very robust against defect scattering when defects
are away from zigzag edges. {Such location-dependent effects can
be depicted by the charge density distribution in Fig. A1. To
reduce defect effect on electron transport, one needs to avoid
creating defects randomly located on the zigzag edges.}

\begin{figure}[h]
\centering
\includegraphics[angle=0,scale=0.3]{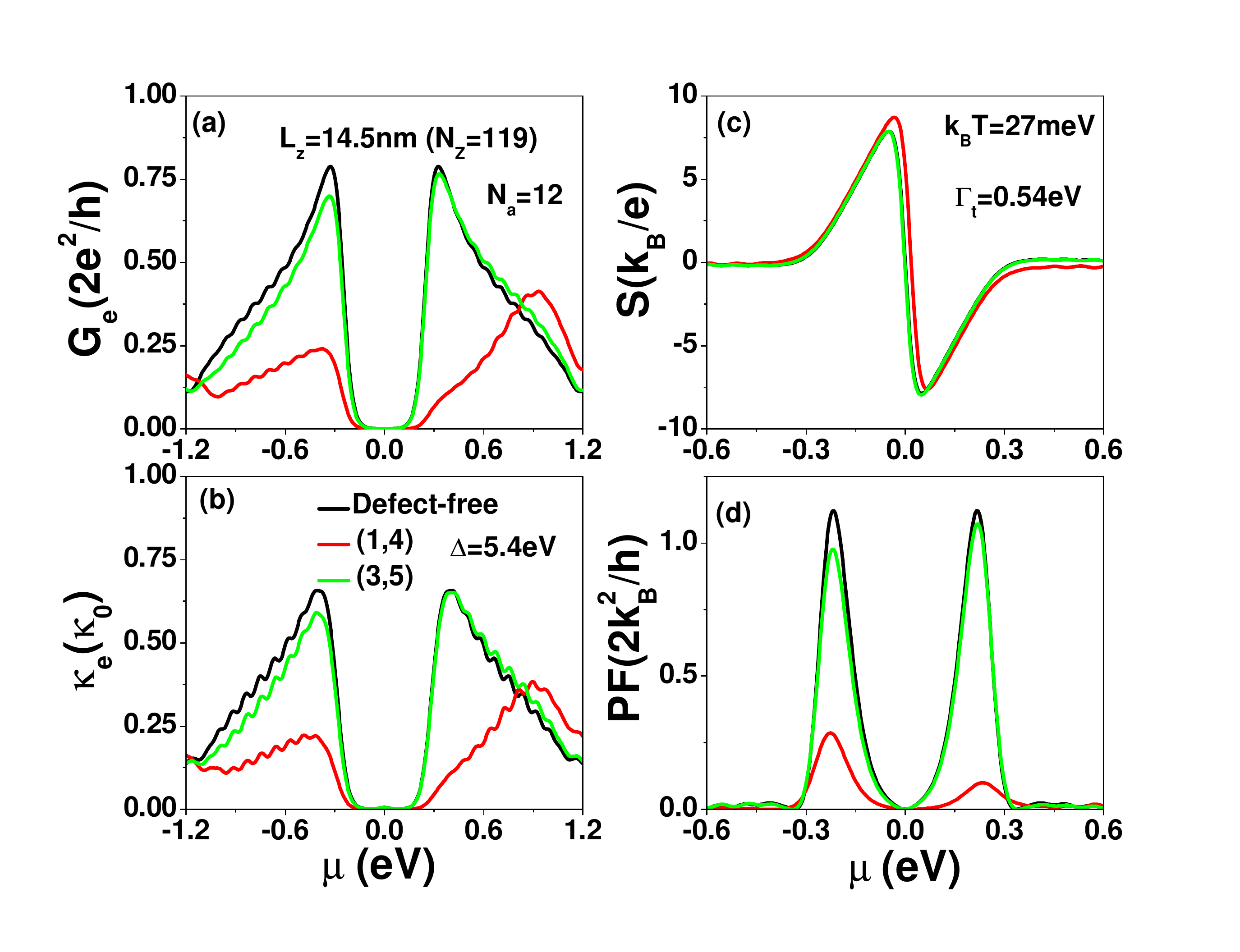}
\caption{ (a) Electrical conductance $G_e$,  (b) electron thermal
conductance $\kappa_e$, (c) Seebeck coefficient $S$, and (d) power
factor$PF=S^2G_e$ as functions of $\mu$ for defect locations at
$k_BT=27meV$.  The tunneling rates used are $\Gamma_t=0.54eV$. The
length of SGQD is $L_z=14.5nm$ ($N_z=119$). Each GQD in the SGQD
structure has size $N_a=12$ and $N_z=7$.}
\end{figure}

We note that SGQDs are formed by periodically removing some carbon
atoms on the zigzag edges of ZGNRs. To understand the relationship
between the quantum states near the central gap in the SGQD formed
by t-ZGNRs and the quantum states near the zero energy modes of
ZGNRs, we show a comparison of subband structures of the infinite
long t-ZGNR shown in Fig. 1(b) and unaltered ZGNR shown in Fig.
1(a) (with an enlarged super cell of length $L=4a$ to match the
unit cell of t-ZGNR) in Fig.~A.2(a) of the appendix. Note that the
first Brillouin zone (BZ) has one quarter the size of the BZ of
ZGNR. Thus, the subband structures shown in Fig.~A.2(a) includes
the zone-folding effect that maps the zone boundary ($k_z=\pi/a$)
of the unfolded BZ of ZGNR to $k_z=0$ of the folded BZ here. As
expected, the unaltered ZGNR has edge states with zero energy for
$k_z <0.06\pi/a=0.24\pi/L$. The two edge states are strongly
coupled in the t-ZGNR with an energy splitting of $~2.5eV$. On the
other hand, the edge states at mini-zone boundary ($k_z=\pi/L$)
with energies corresponding to the HOMO (LUMO) level are nearly
unaffected by the removal of a carbon atom in the neck region of
t-ZGNR, since the wave functions of these edge states have a node
at that position. Theoretical calculations of the {Zak phases} of
various t-ZGNRs indicate that the topological nature of edge
states of t-ZGNRs are preserved even though their energy levels
are shifted away from the CNP [\onlinecite{Chou}]. Thus, these
states are topologically protected and the transport through these
states should be rather insensitive to the presence of defects
inside the t-ZGNR.

Junction tunneling rates can be affected by the quality of contact
between the electrode and semiconductor,[\onlinecite{ChenRS}]
which is a critical issue for device applications of
two-dimensional materials.[\onlinecite{Shen}] To clarify the
contact effect, we show in Fig.~5 the calculated $G_e$,
$\kappa_e$,  $S$, and $ZT$ at room temperature for various
tunneling rates ($\Gamma_t$) {for a t-ZGNR with 15 GQDs, each
having the dimension} $N_a=8$ and $N_z=7$ as shown in Fig. 1(b).
When the tunneling rate increases, the transmission coefficient
${\cal T}_{LR}(\epsilon)$ through minibands close to HOMO (LUMO)
are enhanced. For the case of 15 GQDs, there are 15 peaks in each
miniband (see Fig. A.3). Such an enhancement leads to an increase
of both $G_e$ and $\kappa_e$ as seen in Fig. 5(a) and 5(b).
However, $S$ is essentially independent to the variation of
$\Gamma_t$ as indicated by the collapsing of all four curves in
Fig.~5(c). As a result, the power factor $PF=S^2G_e$ also
increases with $\Gamma_t$. It is noted that the maximum $S$
reaches $1.51mV/K $[see Fig 5(c)], which is much larger than that
observed experimentally in gapless graphene
[\onlinecite{ZuevYM},\onlinecite{WeiP}]. In Fig. 5(d), the maximum
$ZT$ occurs at $\mu=\pm 0.423eV$, where $\kappa_{ph}\gg \kappa_e$.
Therefore, the enhancement of $ZT$ with respect to the increase of
$\Gamma_t$ mainly arises from the enhancement of $PF$. Here, we
have included the effect of phonon thermal conductance,
$\kappa_{ph}$, and assumed $\kappa_{ph}=F_s\kappa^0_{ph}$ in the
calculation of $ZT$ in Fig.~5(d), where
$\kappa^0_{ph}=\frac{\pi^2k^2_BT}{3h}$ is the phonon thermal
conductance of an ideal ZGNR. We adopt the room-temperature value
of $\kappa^0_{ph}=0.285nW/K$ for ZGNR with width $N_a \le 8$
obtained by a first-principles calculation as given in
ref[\onlinecite{Zhengh}]. $F_s=0.1$ denotes a reduction factor
resulting from quantum constriction in t-ZGNRs. It has been
theoretically demonstrated that the magnitude of $\kappa_{ph}$ can
be reduced by one order magnitude for ZGNRs with "edge vacancies"
ref[\onlinecite{Cuniberti}], which is similar to the mechanism
introduced in silicon nanowires with surface
roughness[\onlinecite{Murphy}]. The measured $\lambda_{ph}$ can be
reduced from $300-600nm$ in a single layer graphene to $10nm$ in
graphene nanostructures (see [\onlinecite{XuY}] and references
therein). Recently, very short $\lambda_{ph}$ has been reported
experimentally, which offers promise for enhancing the figure of
merit ($ZT$) of graphene
heterostructures[\onlinecite{WangYH}-\onlinecite{WangYY}].

The calculated results shown in Fig.~5 imply that $ZT > 3$ could
be realized by using SGQDs formed by t-ZGNRs for tunneling rates
corresponding to $\Gamma_t$ between $t_{pp\pi}/10=0.27$ eV and
$t_{pp\pi}/5=0.54$ eV. In Fig.~A.3 of the appendix, we show the
peak value of $ZT (ZT_{max})$ as a function of tunneling rate
$\Gamma_t$, and we found that the $ZT_{max}$ can be larger than 3
for $\Gamma_t$ between $0.18$ eV and $1.45$ eV. We note that if a
semi-infinite ZGNR contact is used, it is reasonable to assume
that the coupling strength between the semi-infinite ZGNR and
t-ZGNR can be comparable to the hopping strength $t_{pp\pi}=2.7
eV$. Here, we show that the maximum $ZT$ can reach 3.7 with
$\Gamma_t =0.54 eV=t_{pp\pi}/5$ that could be achievable by a good
contact.

\begin{figure}[h]
\centering
\includegraphics[angle=0,scale=0.3]{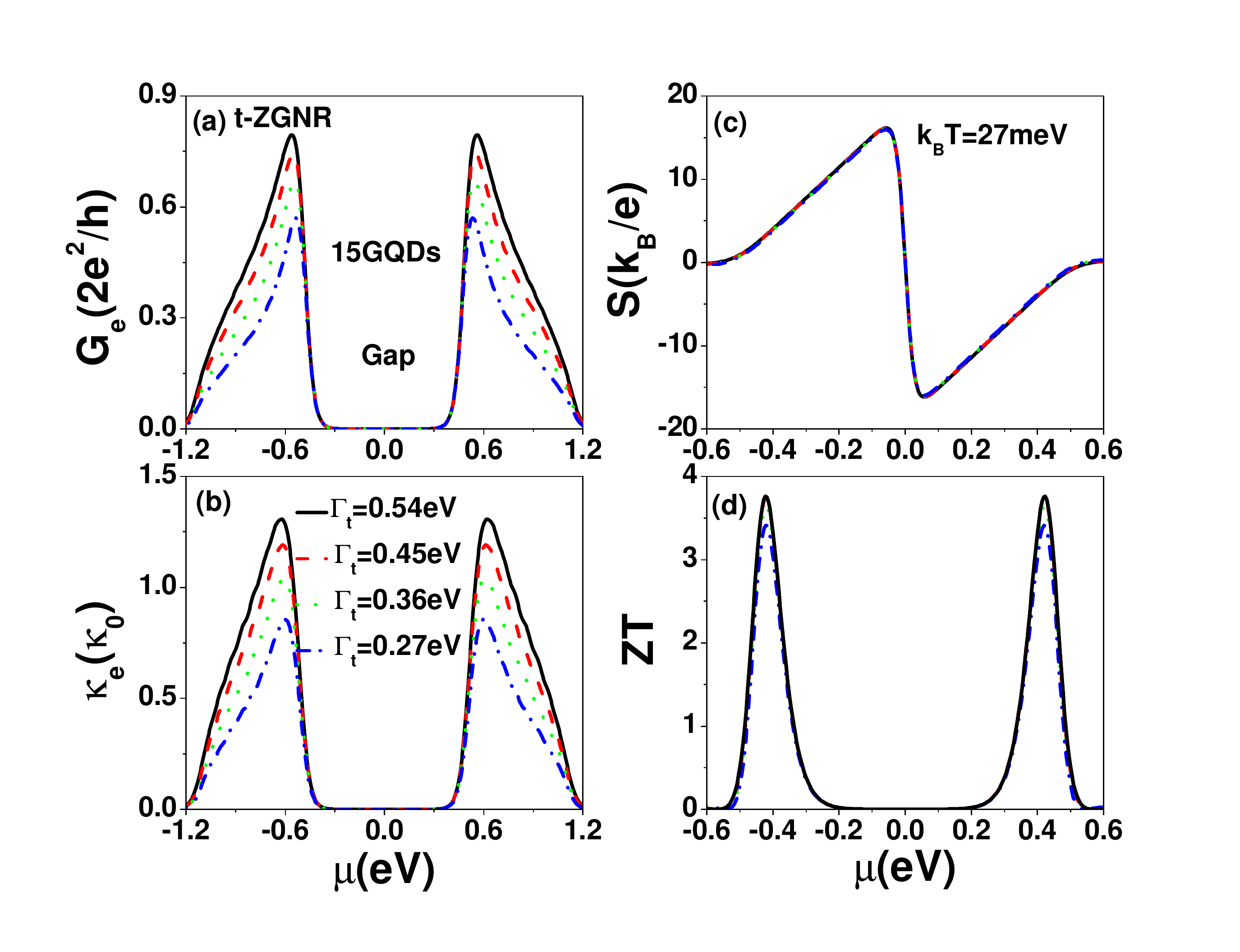}
\caption{(a) Electrical conductance $G_e$,  (b) electron thermal
conductance $\kappa_e$, (c) Seebeck coefficient $S$, and (d)
figure of merit $ZT$ as functions of $\mu$ for various tunneling
rates ($\Gamma_t$) at $k_BT=27meV$.  The t-ZGNR structure with
$L_z=14.5nm$ considered is illustrated in Fig.~1(b). The size of
each GQD in the t-ZGNR is characterized by $N_a=8$ and $N_z=7$.}
\end{figure}

Because $S$ is a robust physical quantity with respect to the
variation of tunneling rate and channel length (which only depend
on the magnitude of central gap and temperature in our case), the
optimization of $G_e$ becomes a critical issue in SGQDs when
$\kappa_{ph}\gg \kappa_e$. To provide better understanding of the
effect of tunneling rate on thermoelectric properties, we further
investigate the relation between tunneling rate and the spectral
shape of ${\cal T}_{LR}(\varepsilon)$. In Fig. 6, we show a
comparison of the tunneling spectra, ${\cal T}_{LR}(\varepsilon)$
{of a t-ZGNR with 15 GQDs} calculated for two different tunneling
strengths $\Gamma_t=0.54eV$ and $\Gamma_t=t_{pp\pi}=2.7eV$. In
Fig.~6(a) the area below the ${\cal T}_{LR}(\varepsilon)$ curve
shows a right-triangle shape that has a steep change with respect
to $\varepsilon$ on the side toward the central gap. In Fig.~6(b)
{(with $\Gamma_t=t_{pp\pi}=2.7eV$) the electron tunneling
probability through the electronic states near
$\varepsilon_{LU(HO)}$ is highly suppressed, leading
to an arch-like area under the ${\cal T}_{LR}(\varepsilon)$ curve.
The corresponding maximum $ZT$ for this case is $~2.3$, which is
much smaller than that of Fig. 5(d). The results of Fig.~6(a) and
Fig.~6(b) indicate that the shape of the area under ${\cal
T}_{LR}(\varepsilon)$ curve depends on the tunneling rate (or the
quality of contact). An arch-like area under ${\cal
T}_{LR}(\varepsilon)$ will reduce the electrical conductance $G_e$
near the central gap.

As seen in Fig.~6(c) the optimized $PF$ found near
$\Gamma_t=0.54eV$ is very close to that obtained by using an ideal
SF transmission spectrum (indicated by the red line), which
exhibits the quantum limit of power factor for 1D systems with
$PF_{QB}=1.2659(\frac{2k^2_B}{h})$
[\onlinecite{Whitney},\onlinecite{ChenIJ}]. We obtain the
optimized $PF_{max}=0.9PF_{QB}$, which is the same as that of the
quantum interference heat engine [\onlinecite{Samuelsson}]. This
analysis suggests that using SGQDs with suitable tunneling rates
could achieve the performance close to an optimum heat engine with
maximum electrical power output and high thermoelectric
efficiency. {With the same $\kappa_{ph}$, $ZT_{max}$ of t-ZGNRs
can reach $95\%$ of that obtained by a SF transmission coefficient
(with} $ZT_{max}=3.951$). So far, we have only considered
periodically indented structures on both the top and bottom zigzag
edges (see Fig. 1(b)). We note that the same optimized result of
$PF_{max}=0.9PF_{QB}$ can also be achieved in periodically
indented structures corresponding to Fig. 6 but with textured
pattern only on one zigzag side provided that the tunneling rate
can be increased to $\Gamma_t=0.72eV$. The results are shown in
Fig.~A.3 in appendix.

\begin{figure}[h]
\centering
\includegraphics[angle=0,scale=0.3]{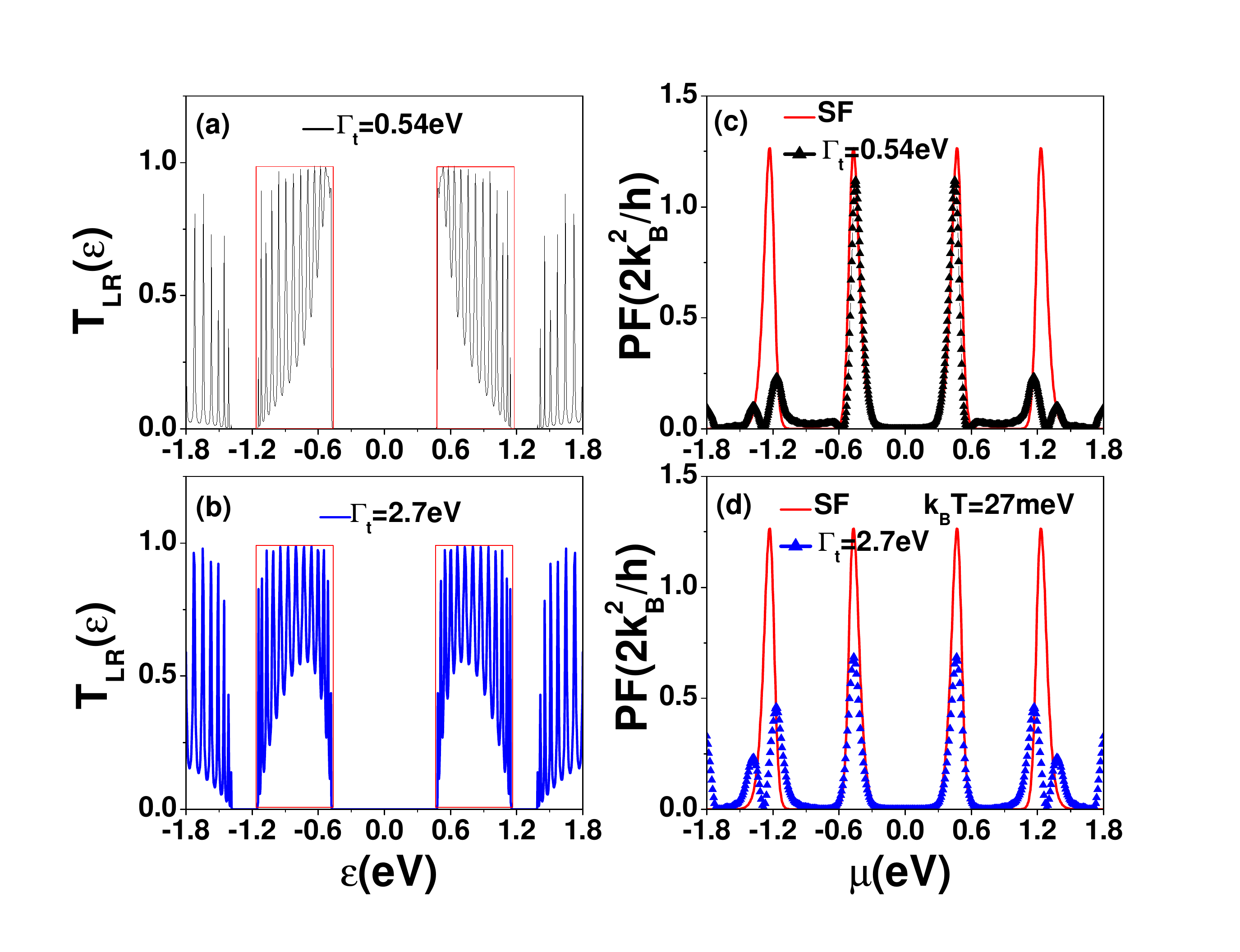}
\caption{(a) and (b) Transmission coefficients ${\cal
T}_{LR}(\varepsilon)$ as functions of energy for a t-ZGNR with $\Gamma_t=0.54eV$
and $\Gamma_t=2.7eV$. The ${\cal T}_{LR}(\varepsilon)$ with square
form (SF) is also plotted by a curve with red color. (c) and (d)
Power factor as functions of $\mu$ for two tunneling rates at
$k_BT=27$ meV. Other physical parameters are the same as those used in
Fig.~5.}
\end{figure}

\subsection{Armchair graphene nanoribbons}
As shown in Figs.~5-6, SGQDs formed by metallic ZGNRs can become
semiconductors. Next, we investigate whether its thermoelectric
performance is better than that of AGNRs or t-AGNRs. We note that
many designs have focused on the optimization of thermoelectric
performance of AGNRs [\onlinecite{Mazzamuto}-\onlinecite{Merino}].
Here, we first consider AGNRs with their zigzag ends coupled to
electrodes as depicted in Fig.~1(c). Figure 7 shows the calculated
transmission coefficient ${\cal T}_{LR}(\epsilon)$ for different
defect locations with $N_z=7$, $N_a=100$ and $\Delta=5.4$eV. We
choose $\Gamma_t=2.7$eV, which gives the optimized shape of the
transmission coefficient in defect free situation in Fig. 7(a).
For finite-size AGNRs the conduction (valence) subband states are
quantized, leading to closely-spaced peaks with staircase-like
structures as revealed by the ${\cal T}_{LR}(\varepsilon)$
spectrum. The area under ${\cal T}_{LR}(\varepsilon)$ curve for
states derived from the first subband has a parabolic shape, which
does not meet the criterium for achieving optimized thermoelectric
property. For a single defect at edge location (1,3), all but the
transmission coefficient spectrum of the first conduction subband
are affected significantly. When the defect is located at site
(3,3), the spectra for the first conduction subband and the first
valence subband are unaffected . Nevertheless, when the defect
occurs at (4,4) the spectra in these two subbands show a
remarkable change.  In current studies we do not observe the
Anderson localization effect, which leads to vanishing $G_e$ in
the subband regions when a single defect occurs on the edges of
ZGNRs and AGNRs [\onlinecite{LeePA}]. This implies that GNRs
considered are not exact 1D systems.

\begin{figure}[h]
\centering
\includegraphics[angle=0,scale=0.3]{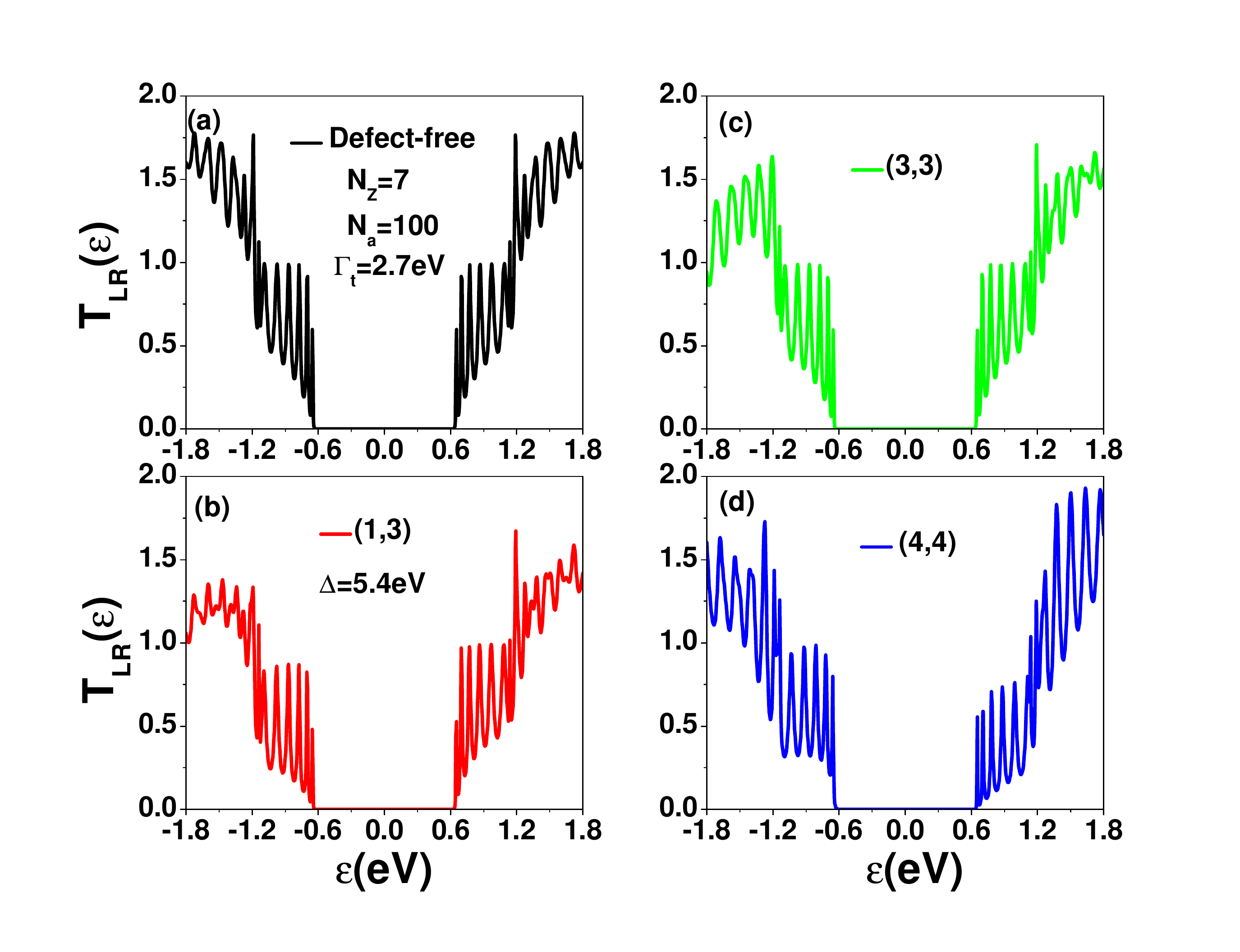}
\caption{Transmission coefficient of AGNRs as functions of
$\varepsilon$ for different defect locations at$\Gamma_t=2.7eV$,
$N_z=7$ and $N_a=100$ ($L_a=10.5$nm). We have adopted
$\Delta=5.4$eV.}
\end{figure}

{To illustrate the effects of defects on thermoelectric properties
of finite AGNRs, we show the calculated electrical conductance
($G_e$), Seebeck coefficient ($S$), power factor ($PF$), and
figure of merit ($ZT$) as functions of $\mu$ at $k_BT=27$ meV in
Fig. 8. Four curves in each diagram correspond to the effects of
defects located at four sites considered in Fig. 7. As seen in
Fig.~8a, the spectra of $G_e$ for defects at locations (1,3) and
(4,4) are degraded seriously either in the valence subband or
conduction subband. Due to the robustness of the Seebeck
coefficient in the central-gap region, the effect of defects on
the power factor is solely determined by $G_e$. Although the
maximum $S$ is larger than that of Fig. 5(c) due to the larger gap
in AGNRs with $N_z=7$, the $PF$ values are smaller than the
corresponding values of the optimized t-ZGNR shown in Fig.~6 (with
$\Gamma_t=0.54$eV).} This is mainly attributed to the different
shapes in the area under the transmission-coefficient curve. As a
consequence, the $G_e$ resulting from thermionic tunneling effect
is less favorable for AGNRs. Note that we have adopted
$\kappa_{ph}=0.0285nW/K$ in Fig. 8(d), which is the same as
$\kappa_{ph}$ in Fig. 5(d).

\begin{figure}[h]
\centering
\includegraphics[angle=0,scale=0.3]{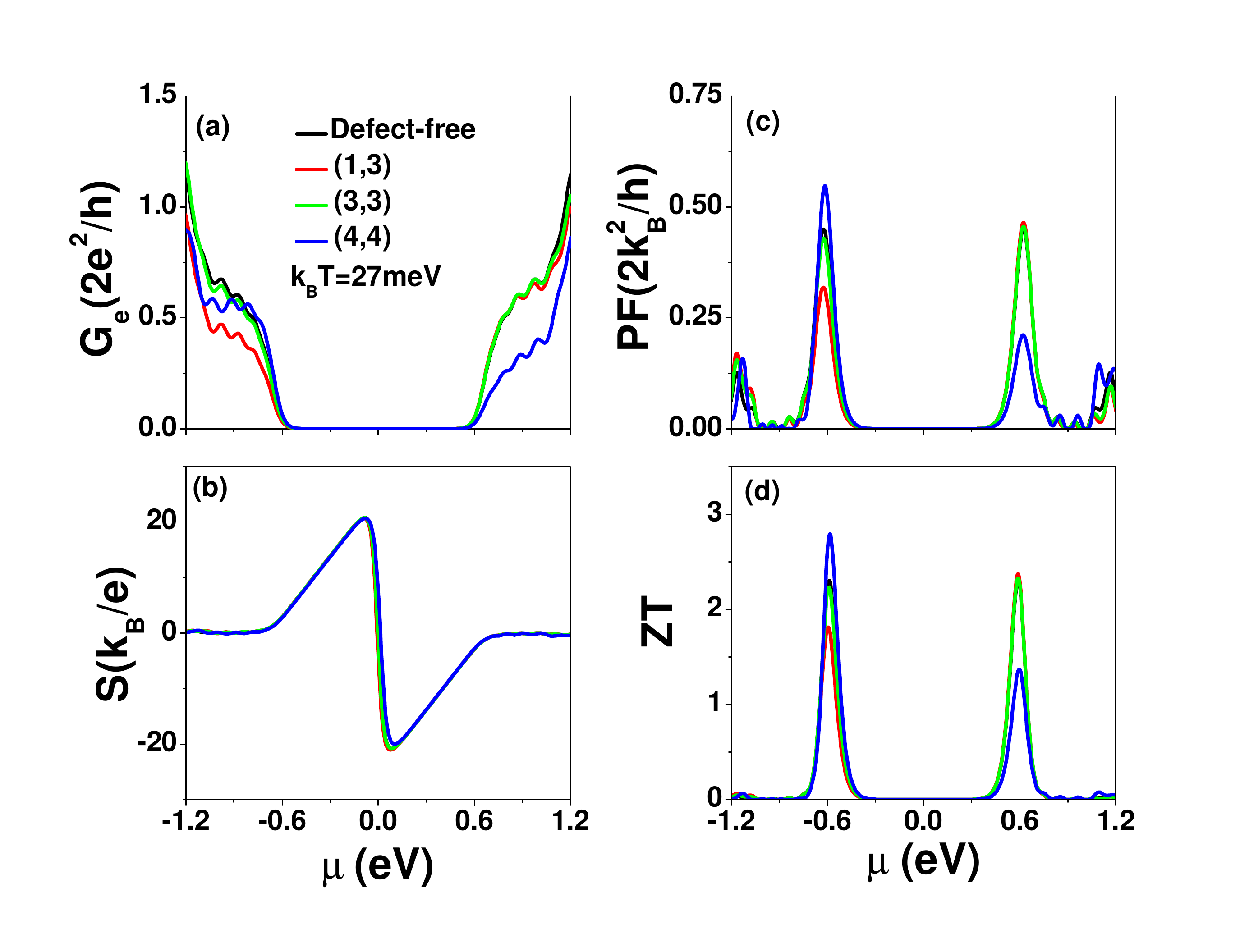}
\caption{(a) Electrical conductance $G_e$, (b) Seebeck coefficient
$S$, (c) power factor $PF$ and (d) figure of merit $ZT$ as
functions of $\mu$ for different defect locations at $k_BT=27meV$.
Other physical parameters are the same as those of Fig. 7. The
considered $\kappa_{ph}=0.0285nW/K$ is the same as that of Fig. 5
($\kappa_{ph}=0.0285nW/K$).}
\end{figure}

\subsection{ SGQDs formed by textured AGNRs}

\begin{figure}[h]
\centering
\includegraphics[angle=0,scale=0.3]{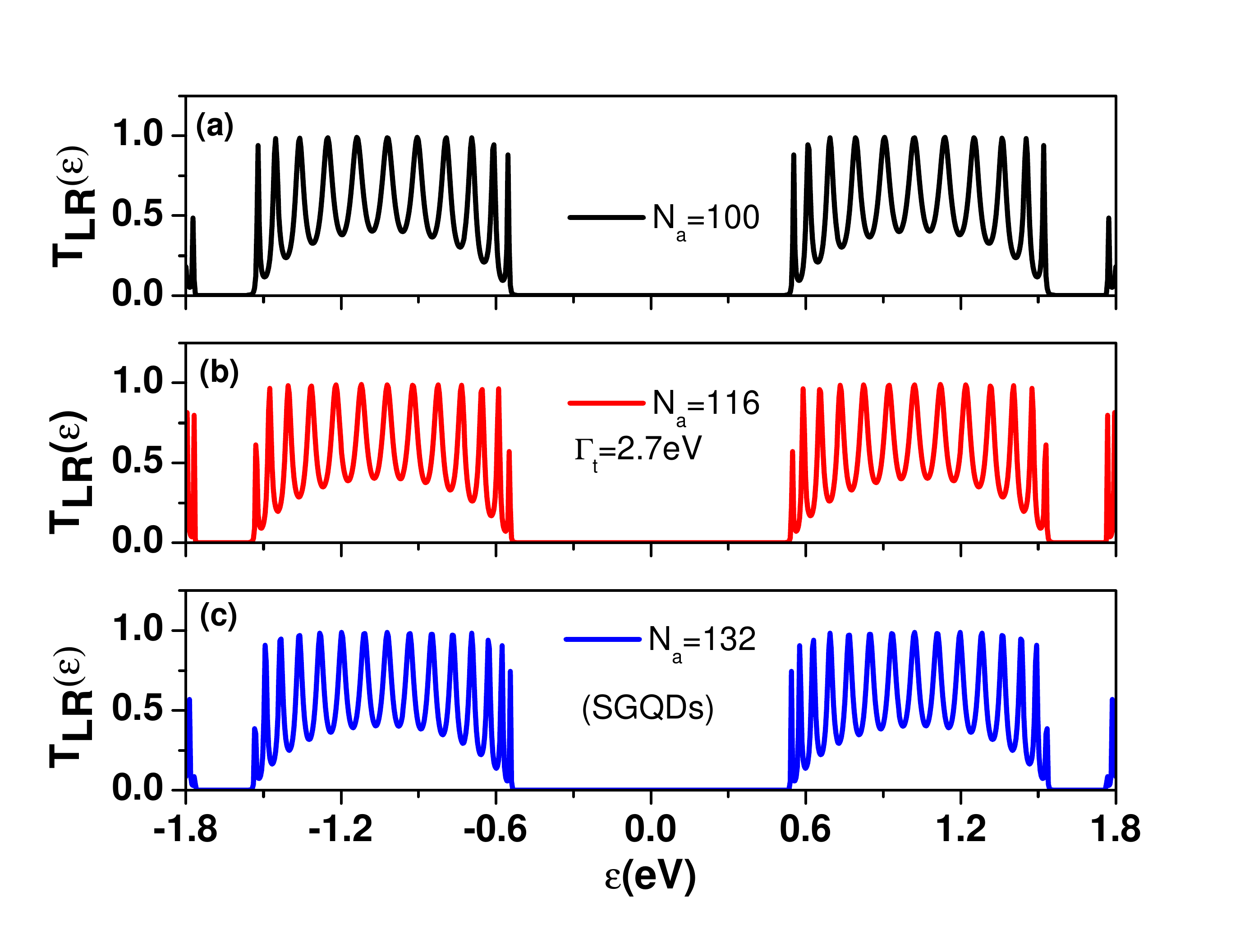}
\caption{Transmission coefficient of SGQD with $N_z=5$ and various
$N_a$ values as functions of energy $\varepsilon$.
$\Gamma_t=2.7eV$. Here, $N_a=12+8\times N_{QD}$, where $N_{QD}$ is
the number of GQDs in the interior region of the structure, and
they are sandwiched between two smaller GQDs of $N_a=6$ connected
to electrodes. Thus, $N_a=100, 116$, and 132 correspond to
$N_{QD}=11, 13$, and 15, respectively. For $N_a=116$ and
$N_a=132$, the lengths of SGQDs are $12.2$ nm and $13.9$ nm,
respectively.}
\end{figure}

Finally, we study the thermoelectric effect of SGQD structures
based on the t-AGNR structure as shown in Fig.~1d. The electronic
structures and density of states (DOS) of t-AGNR superlattices
have been studied theoretically by using density functional theory
(DFT) [\onlinecite{Sevincli},\onlinecite{Rizzo}]. Electronic
structures of t-AGNRs can also be calculated by using a tight
binding model and results are in good agreement with the DFT
calculation[\onlinecite{Sevincli}]. Figure 9 shows the calculated
${\cal T}_{LR}(\varepsilon)$ of SGQD junctions made from t-AGNRs
as functions of $\varepsilon$ for three different $N_a$ values
with $N_z$ fixed at 5. We adopt the optimized tunneling rate with
$\Gamma_t=2.7 eV$. {Although the AGNR with $N_z=5$ is metallic,
the textured AGNR can be semiconducting} as illustrated by the
sizable central gap ($~1.098eV$) in Fig.~9. In Fig.~A.2(b) of the
appendix, we show a comparison of subband structures of the t-AGNR
and unaltered AGNR (with an enlarged super cell of length $L=2a'$
to match the unit cell of t-AGNR), where $a'=\sqrt{3}a$ is the
unit cell length of AGNR. We see that the two subbands with linear
dispersion in the metallic $ N_z=5$ AGNR are split into two
subbands with parabolic dispersion near zero wave-number once the
quantum constriction takes effect in the t-AGNR. It is worth
noting that edge states with zero energy can exist at the left and
right ends (with zigzag edges) of the truncated t-AGNR in contact
with electrodes. These edge states can contribute to ${\cal
T}_{LR}(\varepsilon)$ for t-AGNRs with short SGQD structure (see
Fig. A.4 in appendix), since their wave functions decay
exponentially along the armchair direction. The area under the
${\cal T}_{LR}(\varepsilon)$ curve has an arch-like shape that
does not change much as we vary the tunneling rate.

Figure 10 shows the calculated $G_e$, $S$, $PF$ and $ZT$ of SGQD
junctions made from a t-AGNR with $N_z=5$ and $N_a=132$
($N_{QD}=15$) as functions of temperature for different $\mu$
values. Solid and dashed lines correspond to $\mu=0.52$eV and
$\mu=0.36$eV, respectively. The first conduction subband edge
occurs at $\mu_{edge}=0.544$eV. At a given temperature
$k_BT=27$meV, the maximum power factor occurs at $\mu=0.52$eV. For
$\mu=0.36$eV, which is far away from $\mu_{edge}$, the calculated
$G_e$ in the thermionic assisted tunneling process (TATP) is
extremely small, whereas its Seebeck coefficient is highly
enhanced. The temperature-dependent $S$ for this case is
complicated. Its temperature dependence can be described by three
different functions at three temperature ranges as illustrated in
Fig. 10(b). In region one ($T<90 K$), we have
$S_1=-\frac{\pi^2k^2_BT}{3e} \frac{\partial ln({\cal
T}_{LR}(\varepsilon))}{\partial \varepsilon}|_{\varepsilon=\mu}$.
In region three ($T>120 K$) we have
$S_3=\frac{\mu-\mu_{edge}}{eT}$ [\onlinecite{Kuo1}]. In region 2,
the analytic expression of $S_2$ is unknown. In TATP, $G_e$ can be
described by the expression $exp^{(\mu-\mu_{edge})/(k_BT)}$. As a
consequence, the power factor and figure of merit are small for
$\mu=0.36$eV. When $\mu=0.52$eV, the $PF$ is enhanced quickly in
the range $50 K< T < 120 K$. Within this temperature region, $ZT$
shows a similar behavior of $PF$. When $T > 120 K$}, the
temperature-dependent $ZT$ has the same trend of $S$ because the
heat current is dominated by the linear-$T$ phonon thermal
conductance $\kappa_{ph}=F_s \frac{\pi^2k^2_BT}{3h}$, which
cancels the factor of $TG_e$ in the numerator of $ZT$. It is worth
noting that the optimized $PF$ of t-AGNRs at $T=325K$ is only
one-half of the power factor of the t-ZGNR shown in Fig.~6(c).

Overall, we found no appreciable improvement in the power factor
of t-AGNRs in comparison to AGNRs. Although electron Coulomb
interactions are neglected in this calculation, our conclusion is
still valid even in the Coulomb blockade regime. When the
thermoelectric behavior of SGQDs is dominated by the TATP,
electron-electron correlation functions are usually small
[\onlinecite{Kuo3}]. Therefore, we can neglect electron Coulomb
interactions when $\mu$ is inside the gap between two subbands.

\begin{figure}[h]
\centering
\includegraphics[angle=0,scale=0.3]{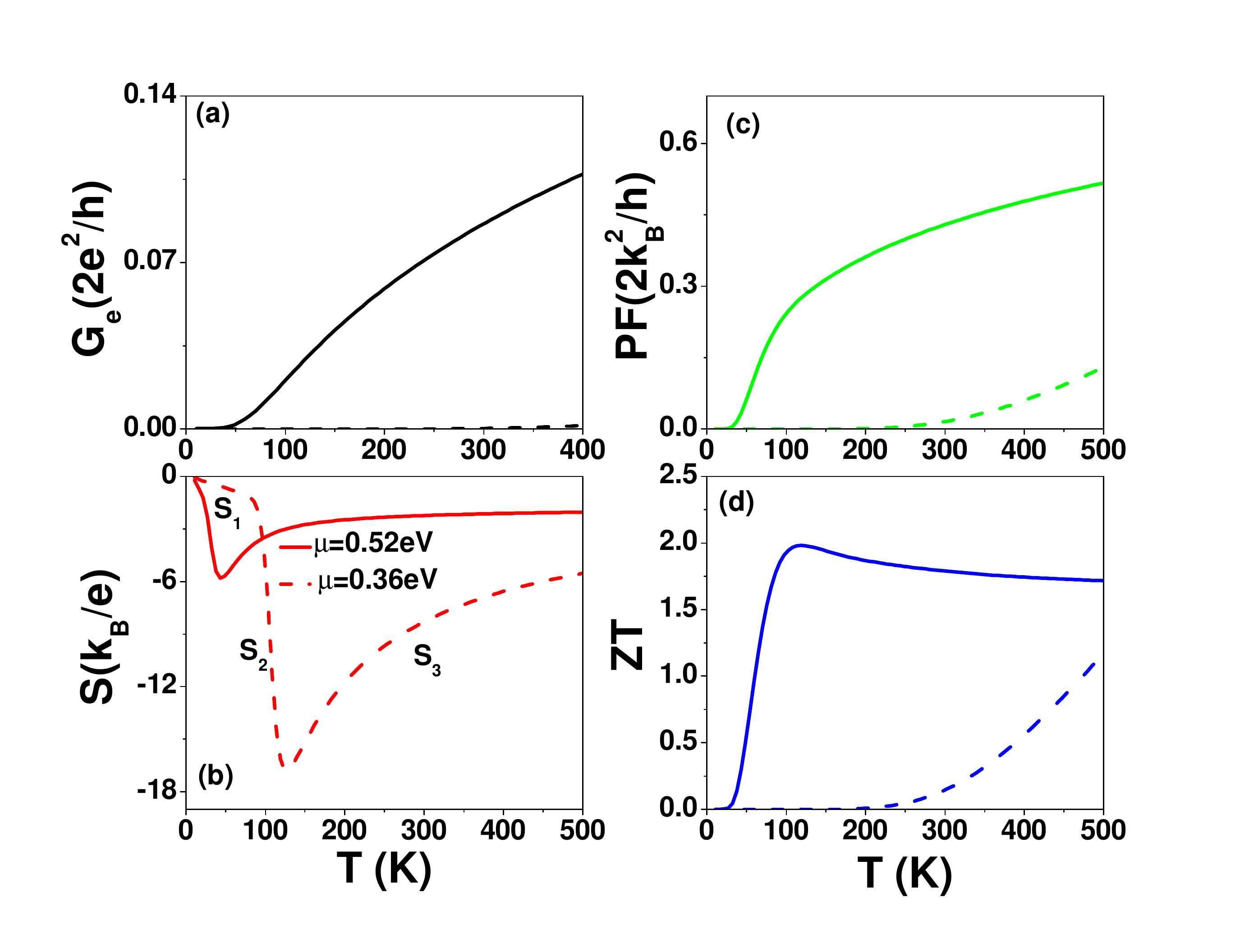}
\caption{(a) Electrical conductance $G_e$, (b) Seebeck coefficient
$S$, (c) power factor $PF$ and (d) figure of merit $ZT$ of SGQD
with $N_z=5$ and $N_a=132$ ($N_{QD}=15$) as functions of
temperature for different $\mu$ at $\Gamma_t=2.7eV$. The phonon
thermal conductance, $\kappa_{ph}$ adopted is the same as in Fig.
8.}
\end{figure}

\section{Conclusion}
We have theoretically investigated the transport and
thermoelectric properties of SGQDs, which are formed by tailoring
ZGNRs and AGNRs. Our calculations are based on the Green's
function approach within a tight-binding model. Electron coherent
tunneling process is found responsible for the electrical
conductance spectra of SGQDs. Subband width and central gap of
SGQDs can be modulated by varying the size of GQDs and inter-dot
coupling strength. Unlike $G_e$ and $\kappa_e$, Seebeck
coefficients are found insensitive to the contact property between
the SGQDs and electrodes. As a result, the power factor and
thermoelectric figure of merit can be improved by modulating the
tunneling rates. The maximum $ZT$ values at room temperature occur
when the chemical potential, $\mu$ is close to the  HOMO (LUMO)
level in typical situation where phonon carriers dominate the heat
transport. {As shown in Fig.~6 and Fig. A.3, the maximum power
factor (figure of merit) of t-ZGNRs at room temperature can reach
90 $\%$ (95 $\%$) of the ideal situation with a square-form
transmission coefficient. We also found that SGQDs based on
t-ZGNRs can outperform SGQDs based on t-AGNRs for thermoelectric
application. The significantly improved thermoelectric behavior of
the textured ZGNR is attributed to the sharp change of its
transmission coefficient near the central gap. We found that
defects at interior sites will not ruin the robust behavior of
{${\cal T}_{LR}(\varepsilon)$ associated with the edge states near
LUMO of t-ZGNRs. It implies that the electron mean free
path,$\lambda_e$ contributed from the zigzag edge states
($\lambda_{e,edge}$) can be much larger than the contribution from
bulk states ($\lambda_{e,bulk}$)). When the channel length ($L_z$)
of t-ZGNRs is much larger than $\lambda_{ph}$ and
$\lambda_{e,bulk}$, $\kappa_{ph}$ is seriously suppressed in such
a diffusing region. The electronic states in the first miniband
(near LUMO) of t-ZGNRs showing nonlinear dispersion could remain
coherent due to their unique nature [\onlinecite{Chou}].} As a
consequence, the power factor of Fig. 6 remains valid, meanwhile
the corresponding $ZT$ could be further
enhanced.[\onlinecite{Xu}]} At room temperature with $T=324K (k_BT
=27meV)$, the electrical power output can reach $0.212nW/K$ for
each SGQD implemented by using t-ZGNR. For an SGQD array with
density of $5\times 10^{6}cm^{-1}$ and $ZT$ larger than 3, the
estimated power output is around $mW/K$, which can be applicable
for low-power wearable electronic devices[\onlinecite{SuarezF}].

%\begin{flushleft}

%\end{flushleft}
{}
%\begin{flushleft}

%{\bf Acknowledgments}\\
%{This work was supported in part by the Ministry of Science and
%Technology (MOST), Taiwan under Contract Nos.
%110-2112-M-001-042 and {110-2119-M-008-006-MBK.} }
%\end{flushleft}

\textbf{Author contributions}\\
{David M T Kuo initiated the idea and performed the calculation.;
David M T Kuo and Y. C. Chang both contributed to the development
of the computation code and carried out the data analyses.David M.
T. Kuo and Y. C. Chang  wrote the manuscript. All authors read and
agreed to the published version of the manuscript}

\textbf{Funding}\\
{This research was funded by the Ministry of Science and
Technology Grant Number MOST 110-2112-M-001-042 and
110-2119-M-008-006-MBK in Taiwan }

\textbf{Data availability}\\
{The data presented in this study are
available upon reasonable request.}

\textbf{Acknowledgments}\\
{This work was supported in part by the
Ministry of Science and Technology (MOST), Taiwan under Contract
Nos. 110-2112-M-001-042 and 110-2119-M-008-006-MBK.}

\textbf{Conflicts of interest}\\
{The authors declare no conflict
of interest.}

%\mbox{}\\
%E-mail address: mtkuo@ee.ncu.edu.tw\\
%E-mail address: yiachang@gate.sinica.edu.tw\\

\appendix
\numberwithin{figure}{section}
\section{}
 \numberwithin{figure}{section} \numberwithin{equation}{section}

\subsection{Charge density of t-ZGNRs}
To further clarify the electronic states shown in Fig.~3, we plot
the charge densities of $|\Psi_{\ell,j}(\varepsilon_{LU})|^2$,
$|\Psi_{\ell,j}(\varepsilon_{e,1})|^2$ and
$|\Psi_{\ell,j}(\varepsilon_{e,2})|^2$ for the {6-GQDs case formed
by t-ZGNRs with $N_a=12$ and $N_z=47$ in Fig.~A.1, which
correspond to the electronic states of $\varepsilon_{LU}=0.247eV$,
$\varepsilon_{e,1}=0.3105eV$ and $\varepsilon_{e,2}=0.468eV$ in
Fig. 3(f). As seen in Fig.~A.1(a), the charge density distribution
in each GQD is the same for the electronic state of
$\varepsilon_{LU}$. The maximum charge densities appearing at
bottom and top zigzag edge sites are contributed by $\ell=1$ and
$\ell=12$ with the condition $j-4=8m$, where $m$ is an integer. At
interior sites of each GQD the charge density is very low. For
$\varepsilon_{e,1}=0.3105eV$, the charge density at the central
two GQDs is negligible. For $\varepsilon_{e,2}=0.468eV$, the
charge density at the $2nd$ and $5th$ GQDs are dilute. For
$\varepsilon_{e,6}=1.0665eV$, the charge density at the $1st$ and
the $6th$ GQDs is smaller than those of other GQDs (not shown
here). This indicates that the electronic states near
$\varepsilon_{e,6}=1.0665eV$ are weakly coupled} to the
electrodes. Based on the charge density distribution shown in Fig.
A.1, the defect location-dependent behavior of $G_e$ shown in Fig.
4 can be clearly illustrated. The $G_e$ does not be degraded
seriously when defects locate at sites with low charge density.

 \begin{figure}[h]
\centering
\includegraphics[angle=0,scale=0.3]{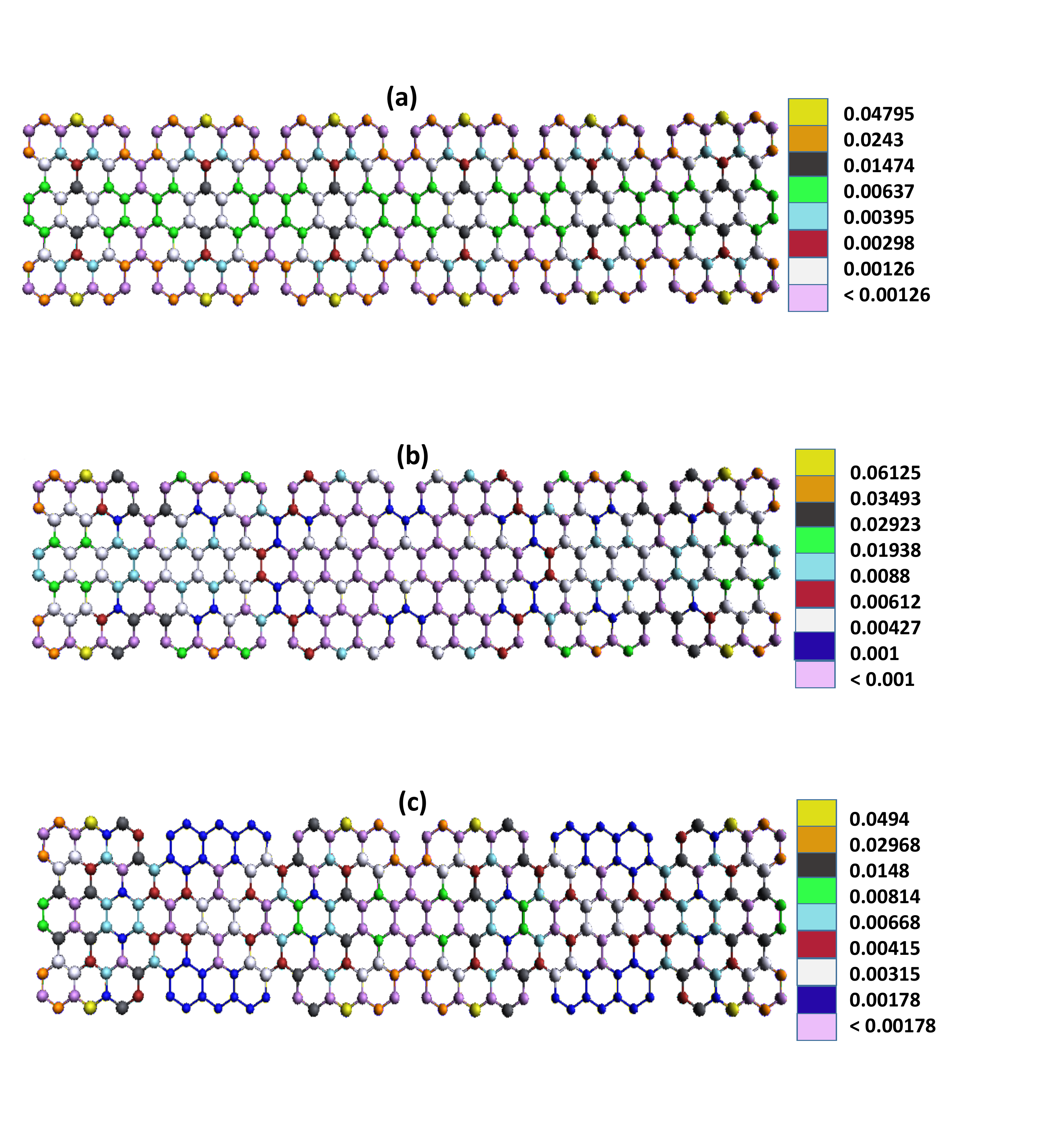}
\caption{Charge density $|\Psi_{\ell,j}(\varepsilon)|^2$ for {the
t-ZGNR with conductance spectrum shown in} Fig. 3(f). Diagrams
(a), (b) and (c) are the charge densities of
$\varepsilon_{LU}=0.247eV$, $\varepsilon_{e,1}=0.3105eV$, and
$\varepsilon_{e,2}=0.468eV$, respectively.}
\end{figure}

\subsection{Electronic band structures}

In Fig.~A.2, we show comparisons of subband structures of graphene
nanoribbons with and without modification of side edges. Here, the
length of unit cell in t-GNRs is denoted $L$. For the t-ZGNR {with
its unit cell characterized by $N_a=8$ and $N_z=8$}, we have
$L=4a$, where $a$ is lattice constant of ZGNR. For the t-AGNR
characterized by $N_z=5$ and $N_a=8$, we have $L=2a'$, where
$a'=\sqrt{3}a$ is the lattice constant of AGNR. We note that the
first Brillouin zone (BZ) has one quarter (half) the size of the
BZ of ZGNR (AGNR). Thus, the subband structures shown in Fig.~ A.1
includes the zone-folding effect that maps the zone boundary
$k_z=\pi/a$ ($k_z=\pi/a'$) of the unfolded BZ of ZGNR (ANGR) to
$k_z=0$ of the folded BZ here. Here, we show the
electronic structure of $N_a=8$ rather than $N_a=12$ in Fig.
A.2(a). For $N_a=8$, we observe not only the central gap, but also
the gap between the first subband and the second subband.

\begin{figure}[h]
\centering
\includegraphics[angle=0,scale=0.3]{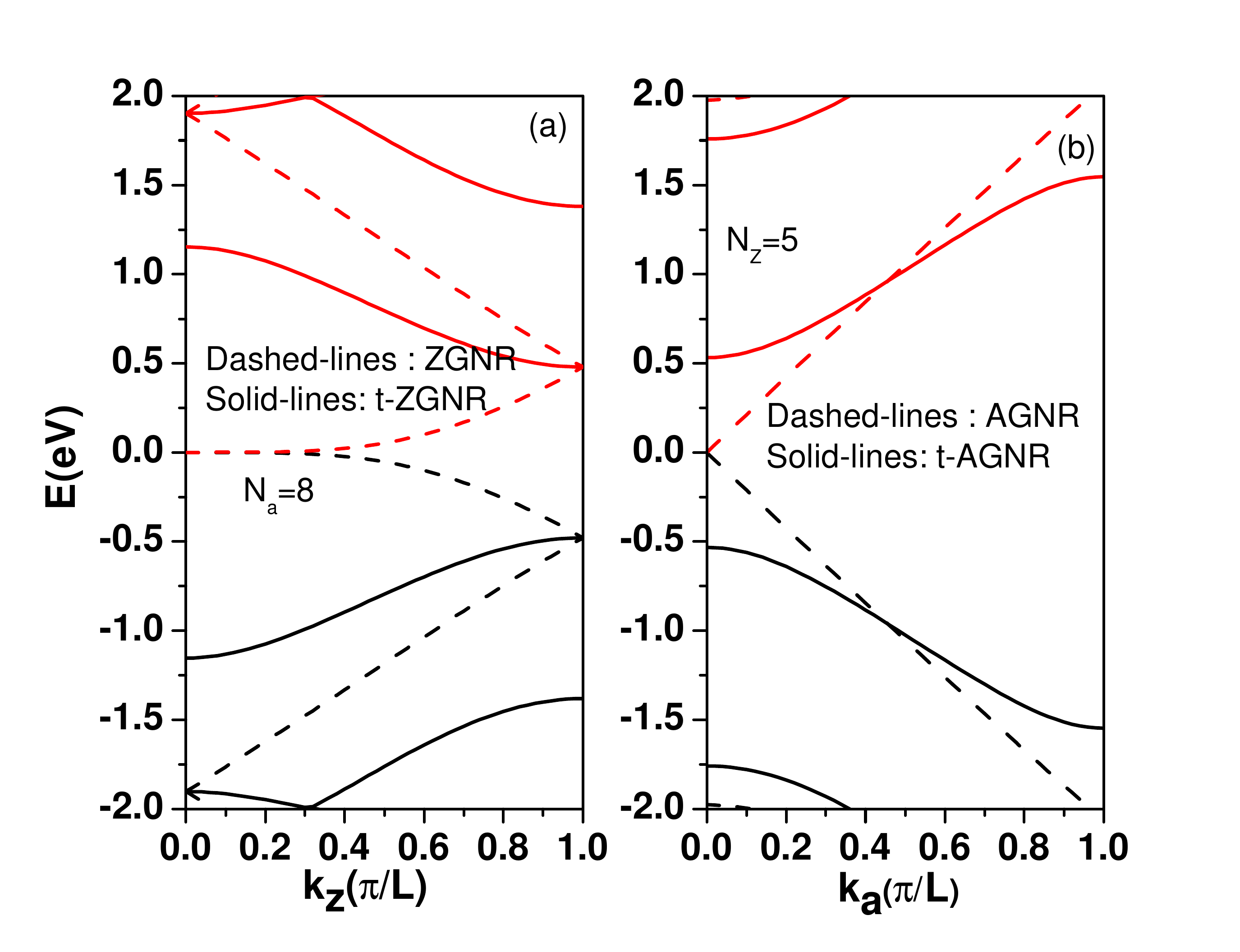}
\caption{(a) Subband structures of t-ZGNR superlattice with  a
unit cell of length $L$ characterized by $N_a=8$ and $N_z=8$
(solid lines) as depicted in Fig.~1b, and corresponding ZGNR
(without texture) of the same length of unit cell (dashed lines).
(b) Subband structures of t-AGNR superlattice with a unit cell
characterized by $N_z=5$ and $N_a=8$ (solid lines) as depicted in
Fig.~1(d), and corresponding AGNR (without texture) of the same
length of unit cell (dashed lines). }
\end{figure}

\subsection{ZT optimization of textured ZGNRs}

In Fig.~A3, we show the calculated ${\cal T}_{LR}(\varepsilon)$ in
the region of $\varepsilon > 0 $ for  {two kinds of t-ZGNRs: (a)
with indentation on both zigzag sides and (b) with indentation on
only one zigzag side of the t-ZGNR. Both t-ZGNRs consist of 15
GQDs} (same as the one considered in Figs.~5 and 6). The solid,
dashed, dotted and dot-dashed lines correspond to $\Gamma_t=0.45,
0.54, 0.63$ and $0.72$eV, respectively. As energy $\varepsilon$ is
tuned away from LUMO, {many sharp peaks appear in the spectrum as
a consequence of the Fabry-Perot interference. The period of
Fabry-Perot interference} in (b) is larger than that in (a). When
$\varepsilon$ gets close to $LUMO$, electronic states can not be
clearly resolved due to their nonlinear dispersion. Diagram (c)
shows $ZT_{max}$ of a t-ZGNR consisting of 15 GQDs as a function
of the tunneling rate $\Gamma_t$ at $k_BT=27$meV and
$\mu=0.423$eV. It is found that $ZT_{max}$ can be larger than 3
for $\Gamma_t$ between 0.18 eV and 1.45 eV for the case of (a).
When the barrier hight gets smaller, the miniband width becomes
wider. This indicates that electron interdot hopping strengths are
enhanced. That is why {$ZT_{max}$ occurs at $\Gamma_t=0.72$ eV in
(b)}. Since the theoretical limit for an ideal SF transmission
coefficient is $ZT_{max}=3.95167$ , the maximum $ZT$ value of
t-ZGNRs can reach $95\%$ of the ideal case.

\begin{figure}[h]
\centering
\includegraphics[angle=0,scale=0.3]{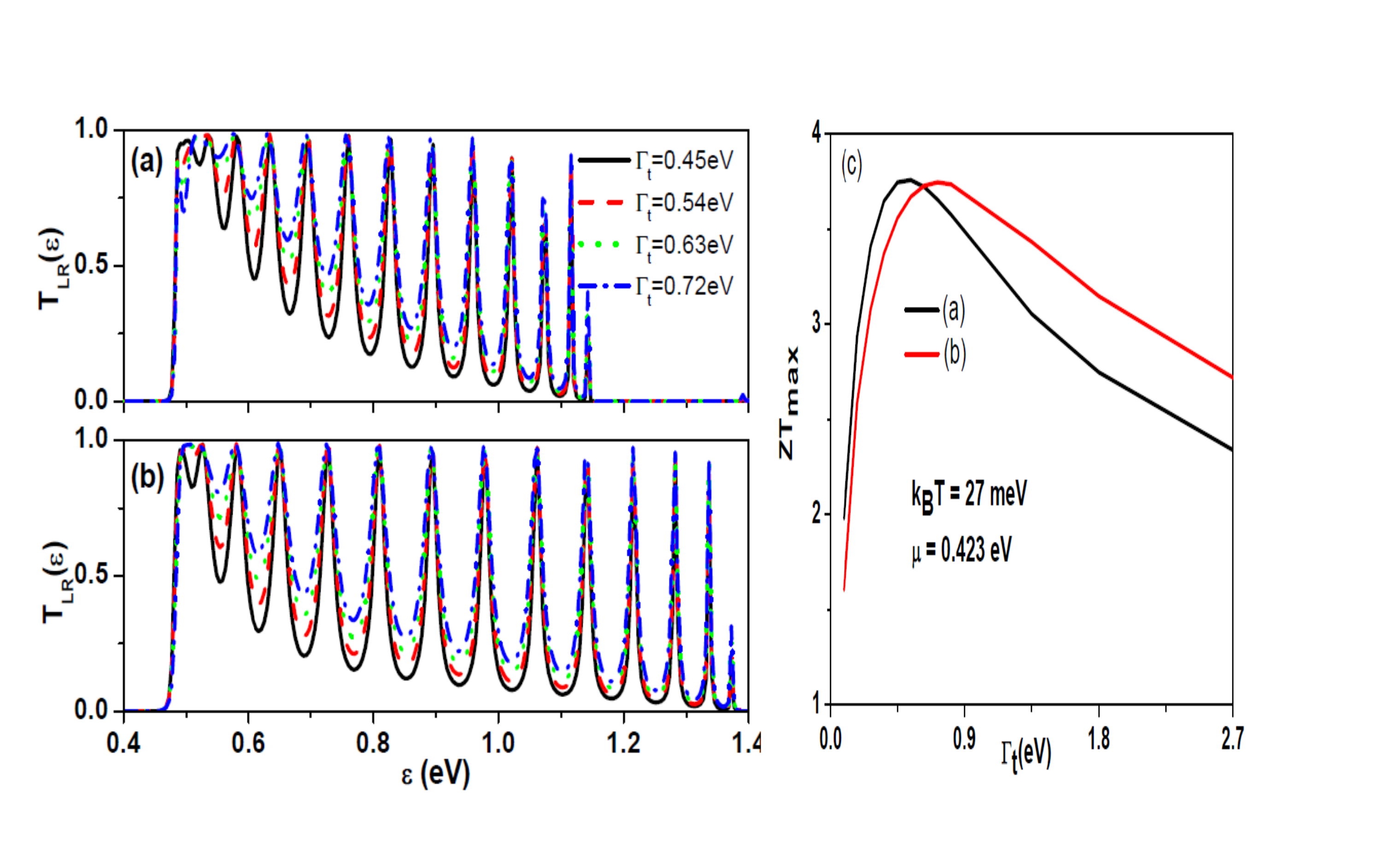}
\caption{Transmission coefficient in the range of $\varepsilon
> 0$ for different tunneling rates. (a) periodically indented structures
on both zigzag-edge sides and (b) periodically indented structures
on only one zigzag-edge side. (c) $ZT_{max}$ of a t-ZGNR
consisting of 15 GQDs as a function of the tunneling rate
$\Gamma_t$. Each GQD in the structure has size $N_a=8$ and
$N_z=7$. The maximum $ZT$ value of SF transmission coefficient at
$k_BT=27$meV is 3.95167 when its $\kappa_{ph}$ is adopted the same
as that of Fig. 5. }
\end{figure}
\subsection{Local density of states of textured AGNRs}
The ${\cal T}_{LR}(\varepsilon)$ {spectrum cannot reveal the
existence of zigzag edge states in a finite-size AGNR that decay
exponentially along the armchair directions when the channel
length is much longer than the decay lengths. These localized
states of GNRs can be probed by STM [\onlinecite{Rizzo}]. In
Fig.~A.4, we show the calculated local density of states (LDOS) as
a function of $\varepsilon$ for various tunneling rates. Here, we
define $LDOS=-Im(G^r_{\ell=2,j=1}(\varepsilon))/\pi$. The LDOS can
reveal the existence of the zigzag edge states of an AGNR coupled
to the electrodes. The zero energy modes resulting from the outer
zigzag edge structures of t-AGNRs are seen in Fig.~A.4. Due to the
localized nature of the zigzag edge states, the spectra in Fig.~9
can not reveal them. Meanwhile, these edge state localized at the
interface are also strongly influenced by the variation of
coupling strength between the electrodes and the t-AGNR. With
increasing $\Gamma_t$, these zero energy modes become broadened}
in (b) and (c).

\begin{figure}[h]
\centering
\includegraphics[angle=0,scale=0.3]{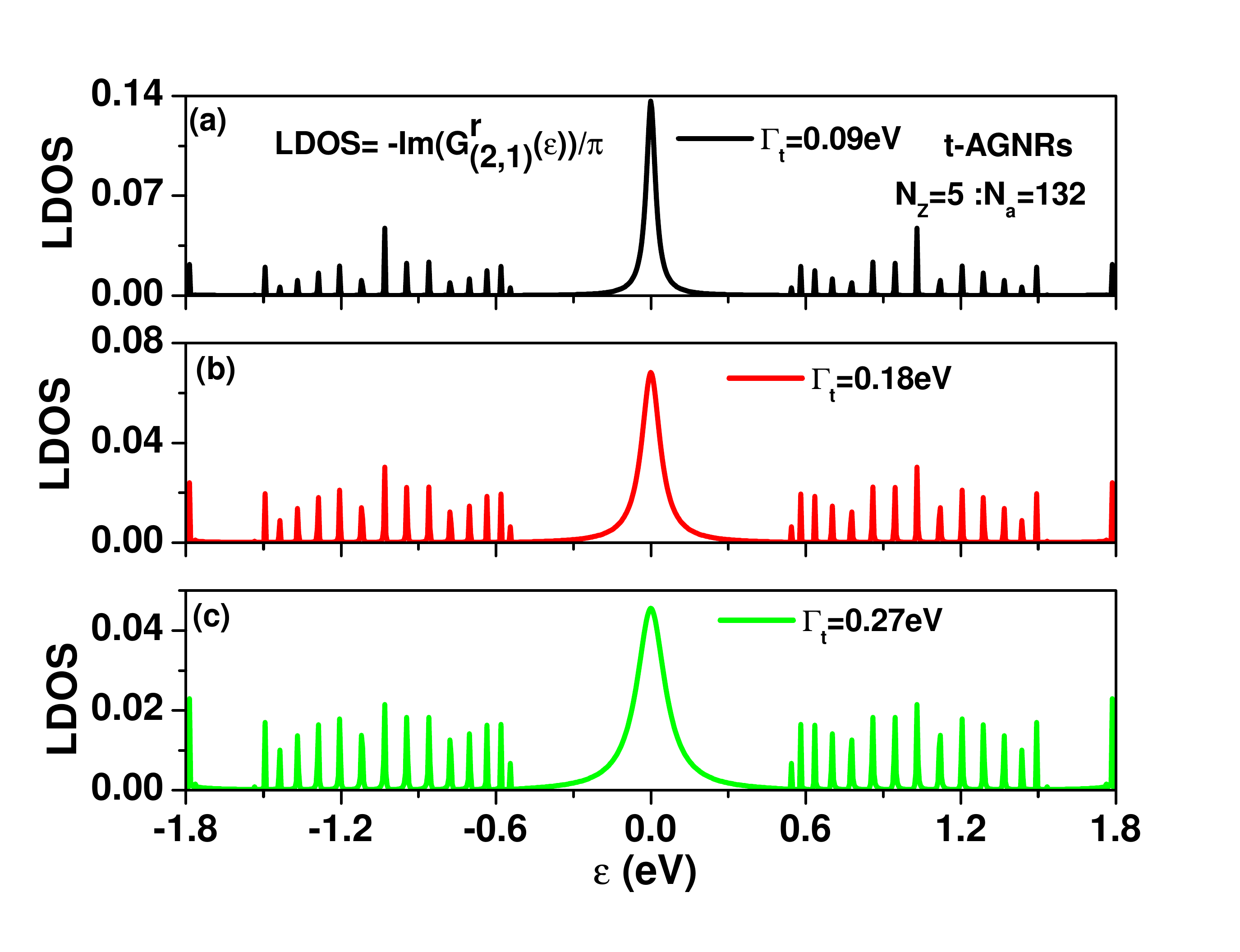}
\caption{Local density of states of {a t-AGNR with
$(N_z,N_a)=(5,132)$ as a function of $\varepsilon$ for various
tunneling rates at site} (2,1).}
\end{figure}

%\section{Appendix}~\Roman{section}
\setcounter{section}{0}
 %\section{Appendix}
\setcounter{equation}{0} % reset counter

%\section{}
%\subsection{Derivation of the tunneling current formula using Dyson's equations\label{App:TC_l} }
\mbox{}\\

%\appendix
%\numberwithin{figure}{section}
%\section{Electronic band structures}

%\numberwithin{equation}{section}

%{\bf Data Availability Statements}\\

%The data that supports the findings of this study are available
%within the article.

\end{document}